\documentclass[aps,twocolumn,prl,superscriptaddress]{revtex4-2}

\usepackage{graphicx}
\usepackage{booktabs}

\usepackage{float} 
\usepackage{mathtools}

\usepackage{amsmath}
\usepackage{yhmath}
\usepackage{amssymb}
\usepackage[export]{adjustbox}
\usepackage{dcolumn}
\usepackage{bm}
\usepackage{hyperref}
\usepackage{enumitem}

\hypersetup{linktocpage,colorlinks,citecolor={blue},pdfdisplaydoctitle=true,pdfpagemode=UseOutlines,bookmarksnumbered=true}

\usepackage{mathrsfs,dsfont}
\usepackage{xcolor}

\usepackage{orcidlink}
\usepackage{bbold}
\usepackage{gensymb}
\usepackage{csquotes}

\usepackage{braket}
\usepackage{cancel}
\usepackage{verbatim}

\graphicspath{{Figures/PNG/}{Figures/PDF/}{Figures/EPS/}{Figures/TEX/}{Figures/}}

\newcommand{\figeq}[2][1cm]{%
  \vcenter{\hbox{\includegraphics[width=#1]{#2}}}%
}

\renewcommand{\L}{\mathcal{L}}

\newcommand{\ssquare}{\mathord{\scalebox{0.5}{$\square$}}}
\newcommand{\blackssquare}{\mathord{\scalebox{0.5}{$\blacksquare$}}}

\newcommand{\refippolitiminimal}{33}
\newcommand{\reffritzschfree}{34}
\newcommand{\reffritzschcumulants}{64}

\begin{document}

\makeatletter
\let\supp@author\author
\let\supp@affiliation\affiliation
\let\supp@email\email
\let\supp@maketitle\maketitle
\let\supp@thanks\thanks
\makeatother

\title{Out-of-Time-Ordered Correlators
Beyond Lindblad }

\author{Elisa Vallini\orcidlink{0009-0003-4613-9526}}
\email{evallini@uni-koeln.de}
\affiliation{Institut f\"ur Theoretische Physik, Universit\"at zu K\"oln, Z\"ulpicher Straße 77, 50937 K\"oln, Germany}

\author{Felix Fritzsch\orcidlink{0000-0002-7659-7574}}
\affiliation{Max Planck Institute for the Physics of Complex Systems, Nöthnitzer Strasse 38, Dresden 01187, Germany}

\author{Pieter W. Claeys\orcidlink{0000-0001-7150-8459}}
\affiliation{Max Planck Institute for the Physics of Complex Systems, Nöthnitzer Strasse 38, Dresden 01187, Germany}
\affiliation{School of Physics, Trinity College Dublin, Dublin 2, Ireland}

\date{\today}

\begin{abstract}
Out-of-time-ordered correlators (OTOCs) provide a powerful probe of quantum chaos and scrambling.
While the Lindblad formalism is often used to describe OTOCs in open systems, microscopic mechanisms that give rise to the Lindblad equation at the level of time-ordered correlation functions do not necessarily justify this approach at the level of OTOCs.
Starting from a microscopic model of a subsystem coupled to a random bath, we show that the resulting OTOC dynamics contains an additional contribution compared to the standard Lindblad approach.
This contribution, induced by correlations between different replicas, allows key properties of the underlying unitary dynamics such as long-time freeness to be retained and information scrambling to be captured more faithfully. 
As an application, we show that this extended formalism can be used to numerically study OTOCs in large closed systems by replacing part of the system with a dissipative boundary, offering a way to study operator spreading and higher-order correlations beyond directly accessible system sizes.
\end{abstract}

\maketitle

\emph{Introduction.} ---
Quantum information scrambling has emerged as a fundamental characterization of chaos in interacting many-body systems over the last decade. Out-of-time-ordered correlators (OTOCs), originally introduced in the context of superconductivity~\cite{larkin1969quasiclassical}, provide a powerful probe of scrambling by quantifying operator growth and the spreading of quantum information~\cite{garcia2022out,hosur2016chaos,maldacena2016bound,xu2024scrambling, tsuji2018bound, trunin2023refined, nahum2018operator}, and have become experimentally accessible in quantum simulators~\cite{li2017measuring}.
Attention has recently turned to the study of OTOCs in open quantum systems, revealing how dissipation competes with unitary operator growth and suppresses information scrambling~\cite{poulin2010lieb, carolan2024operator,jiang2025refining,schuster2023operator,srivatsa2024operator,zanardi2021scrambling, bhattacharjee2023operator, lovas2024coding}.
Despite this progress, a fundamental challenge remains. While Markovian reduced dynamics is described by the Gorini--Kossakowski--Sudarshan--Lindblad (GKSL) equation for single operators and conventional time-ordered correlation functions~\cite{gorini1976completely,lax1963formal,lindblad1976generators,swain1981master}, its extension to higher-order correlators is more subtle. Since an OTOC involves multiple replicas of the same time-evolved operator, these replicas cannot in general be evolved independently under the reduced dynamics~\cite{syzranov2018otoc, blocher2019regression, tripathy2026speed, zhai2020otoc}.

In this work, we contrast $\langle A(t)BA(t)B\rangle_{\L}$, the ``full'' OTOC between observables undergoing full unitary evolution under joint system-bath dynamics, with $\langle A_{\L}(t)BA_{\L}(t)B\rangle$, the ``reduced'' OTOC for which the observables undergo independent Lindblad dynamics. As our main result, we develop a systematic description of OTOC evolution in open quantum systems governed by Lindblad dynamics, keeping track of correlations between replicas of the environment present in the full OTOC. These correlations, originating from the multiple copies of the time-evolution operator, encode information beyond the reduced Lindblad OTOC and lead to a nonperturbative correction that can nevertheless be expressed purely in terms of Lindblad dynamics. Our results are derived by taking the continuous-time limit of an exactly solvable minimal quantum circuit model giving rise to Markovian dynamics.

We apply this formalism to a chaotic spin chain with local boundary dissipation, showing that the resulting OTOC provides a faithful description of scrambling. The replica-correlation term restores fundamental properties inherited from the underlying unitary evolution, including conservation of the operator Hilbert--Schmidt norm, also referred to as conservation of quantum information, and the correct late-time behavior of the closed system, both of which are lost in the conventional Lindblad formalism. The OTOC within this extended formalism therefore more accurately approximates that of the corresponding closed system, suggesting a practical route for simulating scrambling in large isolated systems through suitably engineered open-system simulations. Related strategies have been successfully employed to study correlation functions relevant, e.g., for hydrodynamics~\cite{elbracht2020accessing,vonKeyserlingk2022operator,rakovsky2022DAOE,heitmann2023spin,kraft2024lindblad,kuo2024energy,srivatsa2026hydro}; our formalism extends this perspective to higher-order correlation functions.

\emph{Model.} --- 
The starting point of our approach is the minimal quantum circuit model from Refs.~\cite{ippoliti2022solvable,fritzsch2026free}, which allows for an exact solution of the OTOC dynamics.
The model mimics a system $A$ coupled to a maximally scrambling environment $E$ via an intermediate site $C$. The unitary dynamics over a single discrete time step $\Delta t$ is given by a unitary evolution operator $\mathcal{U}_{\Delta t} = (\mathbb{1}_{A} \otimes V_{CE})(U_{AC}\otimes \mathbb{1}_{E})$. The scrambling dynamics is modeled by a Haar-random unitary $V_{CE}$, chosen independently at every time step, and the structured dynamics is mediated by a fixed gate $U_{AC} = U_{\Delta t}$.

The bath $E$ thermalizes $C$ after each random interaction, such that $A$ effectively interacts with a freshly prepared site $C$ at every time step. The model can therefore be viewed as a repeated-interaction model~\cite{attal2006repeated,strasberg2017quantum,ciccarello2022collision}. 
Anticipating a continuum time limit, we parameterize the unitary gate as $U_{\Delta t} = \exp(-i \tilde{H}_{AC})$ with
\begin{equation}
    \tilde{H}_{AC} = \Delta t 
\left(H_A \otimes \mathbb{1}_C
\!+\! \mathbb{1}_A \otimes H_C\right)+\sqrt{\Delta t}
\sum_\alpha J^{(\alpha)}_{A}\!\otimes J_{C}^{(\alpha)},
\label{Ucont_time}
\end{equation}
with Hermitian and traceless operators $H_A$ and $H_C$ and interaction terms $J_{A}^{(\alpha)}$ and $J_{C}^{(\alpha)}$~\footnote{For convenience, we choose the operators $J_C^{(\alpha)}$ to be orthonormal with respect to the normalized Hilbert--Schmidt inner product, so that $J_A^{(\alpha)}$ directly become the Lindblad jump operators after the continuous-time limit. This assumption can be relaxed by absorbing the resulting correlation matrix into a redefinition of the jump operators.}.
The absence of $\Delta t$ in the scrambling dynamics $V_{CE}$ effectively induces a separation of time scales.
Our first aim is to show that this circuit reproduces effective Lindblad dynamics for single operators of the subsystem $A$, upon averaging over the random environment and taking the thermodynamic and continuous-time limits. We then show that applying the same derivation to the OTOC gives rise to an additional contribution.

\emph{Correlation functions and OTOCs.} --- 
We now consider the dynamics of time-ordered and out-of-time-ordered correlators between observables $A_A$ and $B_A$, which act on the full Hilbert space but are supported only on subsystem $A$: $A$ and $B$ denote their nontrivial parts. In the limit of an infinite environment and upon averaging over the random unitaries, their correlation functions and OTOCs are  
\begin{align}
    \langle A(t)B\rangle_{\L} & \equiv \lim_{d_E\rightarrow \infty} \mathbb{E}[\,\langle A_A(t)B_A \rangle\,]\ , 
    \label{C1_full} \\
    \langle A(t)BA(t)B\rangle_{\L} &\equiv \lim_{d_E\rightarrow \infty} \mathbb{E}[\,\langle A_A(t)B_AA_A(t)B_A \rangle\,] \ ,
    \label{C2_full}
\end{align}
defined with respect to the infinite temperature state, $\langle\bullet\rangle = \text{Tr}(\bullet)/\text{Tr}(\mathbb{1})$. Both the average and the thermodynamic limit can be carried out explicitly~\cite{fritzsch2026free}, see also~\cite{SuppMat2026} for details.
In Eq.~\eqref{C1_full}, at the level of correlation functions the evolution of $A$ can be described in terms of a reduced unital CPTP map.
The unitary gate~\eqref{Ucont_time} is parameterized such that, for small $\Delta t$, this map becomes~\cite{SuppMat2026}
\begin{align}
   \frac{1}{d_C}\mathrm{Tr}_C\!\left[U_{\Delta t}(A\otimes\mathbb{1}_C)U_{\Delta t}^\dagger\right]
   = A+\Delta t\,\mathcal{L}(A)+\mathcal{O}(\Delta t^{3/2}) \ ,
\label{ev_1op}
\end{align}
with Lindbladian
\begin{equation}
    \mathcal{L}(A)
    =-i[H_A,A]
    +\sum_\alpha\left(J_\alpha A J_\alpha-\frac{1}{2}\{J_\alpha^2,A\}\right),
\label{Lindblad_operator}
\end{equation}
where $J_\alpha=J_A^{(\alpha)}$ are Hermitian jump operators.
In the continuous-time limit the resulting correlation function reads
\begin{align}
    \langle A(t)B\rangle_{\L} = \langle A_{{\L}}(t)B\rangle \ ,
\label{eigenop_1}
\end{align}
where the operator $A$ evolves according to the Lindblad equation,
\begin{equation}
   \partial_t A_{\L} = \mathcal{L}(A_{\L}) \quad \implies \quad  A_{\L}(t)=e^{t\mathcal{L}}(A) \ .
\end{equation}
As a first result, we have shown that the random circuit reproduces Lindblad dynamics in a controlled manner. While conventional correlation functions are fully captured by this Lindblad description, OTOCs are not.
In the continuous-time limit the OTOC reads~\cite{SuppMat2026}
\begin{align}
     &\langle A(t)BA(t)B\rangle_{\L} =  \langle A_{\L}(t)BA_{\L}(t)B\rangle \, \nonumber \\ 
     &\qquad \qquad  -\sum_\alpha \int_0^t ds \, \langle \left([A_{\L}(s), J_\alpha] B_{\L^\dagger}(t-s)\right)^2\,\rangle \ .
\label{eq:OTOC_full_main}
\end{align}
We again have that $A_{\L}(t)=e^{t\mathcal{L}}(A)$, while $B_{\L^\dagger}(t)=e^{t\mathcal{L}^\dagger}(B)$ evolves under the adjoint Lindblad generator.
The first term coincides with the OTOC obtained by independently evolving the two copies of $A(t)$ with the usual Lindbladian.
The second term, however, originates from correlations between the two copies of the full unitary evolution.
We denote this correction by $I_{AB}(t)$. 
As the integrand depends on all intermediate times, it is tempting to relate this correction to memory effects in the bath. This is, however, not the case. 
The correction originates purely from correlations between different replicas of the environment, which itself is perfectly Markovian.
Intuitively, the integrand describes an interaction between the two replicas at an intermediate time, corresponding in the folded picture to a jump between different replica pairings.

\emph{Conservation of quantum information.} --- While unitary dynamics preserves the Hilbert--Schmidt norm of $A(t)$, also known as conservation of quantum information~\cite{xu2024scrambling}, Lindblad dynamics generally leads to a decaying norm and information loss~\cite{schuster2023operator}. The replica-correlation term can be seen as restoring this conservation law. Choosing $B=\mathbb{1}$, $\braket{A(t)A(t)} = \braket{A^2}$ is constant in time for unitary evolution. Conversely, within standard Lindblad dynamics, $\braket{A_{\L}(t)A_{\L}(t) }$ generally decays exponentially in time.
A direct calculation gives
\begin{align}
    \frac{d}{dt}\braket{A_{\L}(t)A_{\L}(t)} =& \sum_\alpha \langle [A_{\L}(t),J_\alpha]^2\rangle  = -\frac{d}{dt} I_{A,B=\mathbb{1}}(t) \ ,
    \label{eq:decay_A2_vs_I}
\end{align} 
from which it directly follows that
\begin{align}
    \frac{d}{dt} &\langle A(t)A(t)\rangle_{\L}  = 0  \quad \Longrightarrow \quad \langle A(t)A(t)\rangle_{\L}   = \langle A^2\rangle \ .
    \label{property}
\end{align}
The conservation of the Hilbert--Schmidt norm is in fact a particular instance of a more general property: the replica correction restores the multiplicative structure of the underlying unitary dynamics, which is lost in the reduced Lindblad description~\cite{SuppMat2026}.

\emph{Illustration using a spin chain.} --- We now illustrate the OTOC in a one-dimensional spin chain. Here, we focus on operator spreading in many-body models, where the locality of the interactions and conservation of quantum information lead to a hydrodynamic description. A single-spin example is discussed in~\cite{SuppMat2026}. We consider a spin-$1/2$ Ising chain with mixed fields and local dissipation at the right boundary,
\begin{equation}
    H = J\sum_{i=1}^{L-1}\sigma_i^z \sigma_{i+1}^z
    + h_x \sum_{i=1}^L \sigma_i^x
    + h_z \sum_{i=1}^L \sigma_i^z \ ,
    \label{chain}
\end{equation}
with the single jump operator $J=\sqrt{\gamma}\sigma_L^z$. The mixed-field Ising chain is a paradigmatic nonintegrable model exhibiting
quantum-chaotic behavior~\cite{pappalardi2025full}. We fix $J=1$, $h_x=\sqrt{5}/2$ and $h_z=(\sqrt{5}+5)/8$.
 
The OTOC probes operator spreading and scrambling: under unitary dynamics, an initially local operator develops support over increasingly distant sites, giving rise to a characteristic light cone. The propagation of the operator front is characterized by a butterfly velocity, bounded by the Lieb--Robinson velocity~\cite{lieb1972finite}.
How the presence of dissipation modifies operator spreading has been intensively studied~\cite{carolan2024operator,jiang2025refining,schuster2023operator,srivatsa2024operator,zanardi2021scrambling, bhattacharjee2023operator, lovas2024coding, poulin2010lieb}.
The model of Eq.~\eqref{chain} is particularly suitable for studying the competition between unitary and dissipative effects, as the dissipation is restricted to a single boundary. 

We now investigate the role of the replica-correlation term $I_{AB}(t)$ in information scrambling. In Fig.~\ref{Scrambling} we compare the full and the reduced OTOCs with increasing dissipation strength $\gamma$, for $A=\sigma^z_i$, where $i$ ranges from $1$ to $L$, and $B=\sigma_1^z$, located at the left boundary, i.e., the site farthest from the dissipation~\footnote{The same qualitative behavior is observed for jump operators that do not commute with the observable, showing that our conclusions do not rely on this particular choice.}. For $\gamma=0$, both OTOCs coincide and display the expected light-cone spreading from the left boundary into the bulk. As $\gamma$ increases, the two OTOCs start to differ, particularly near the dissipative boundary.
The full OTOC remains qualitatively similar to the unitary dynamics, with dissipation predominantly affecting sites near the dissipative boundary and modifying their long-time behavior. In addition, the propagation of the operator front is slightly slowed down.
By contrast, the reduced OTOC exhibits an earlier decay near the dissipative boundary, whose onset shifts to progressively earlier times as the dissipation strength is increased.
This behavior has a simple interpretation in terms of the decay of the Hilbert--Schmidt norm under the reduced Lindblad dynamics. As $A_{\mathcal L}(t)$ spreads towards the dissipative boundary, its norm starts to decrease, causing the reduced OTOC to decay even before $A_{\mathcal L}(t)$ reaches the support of $B$. Indeed, while the two operators remain spatially separated,
$
    \langle A_{\mathcal L}(t) B A_{\mathcal L}(t) B\rangle
    \approx
    \langle A_{\mathcal L}(t)A_{\mathcal L}(t)\rangle
    \langle B^2\rangle .
$
The replica-correlation term prevents this spurious decay in the full OTOC, consistent with Eq.~\eqref{property}.

\begin{figure}[t]
\centering
\includegraphics[width=\linewidth]{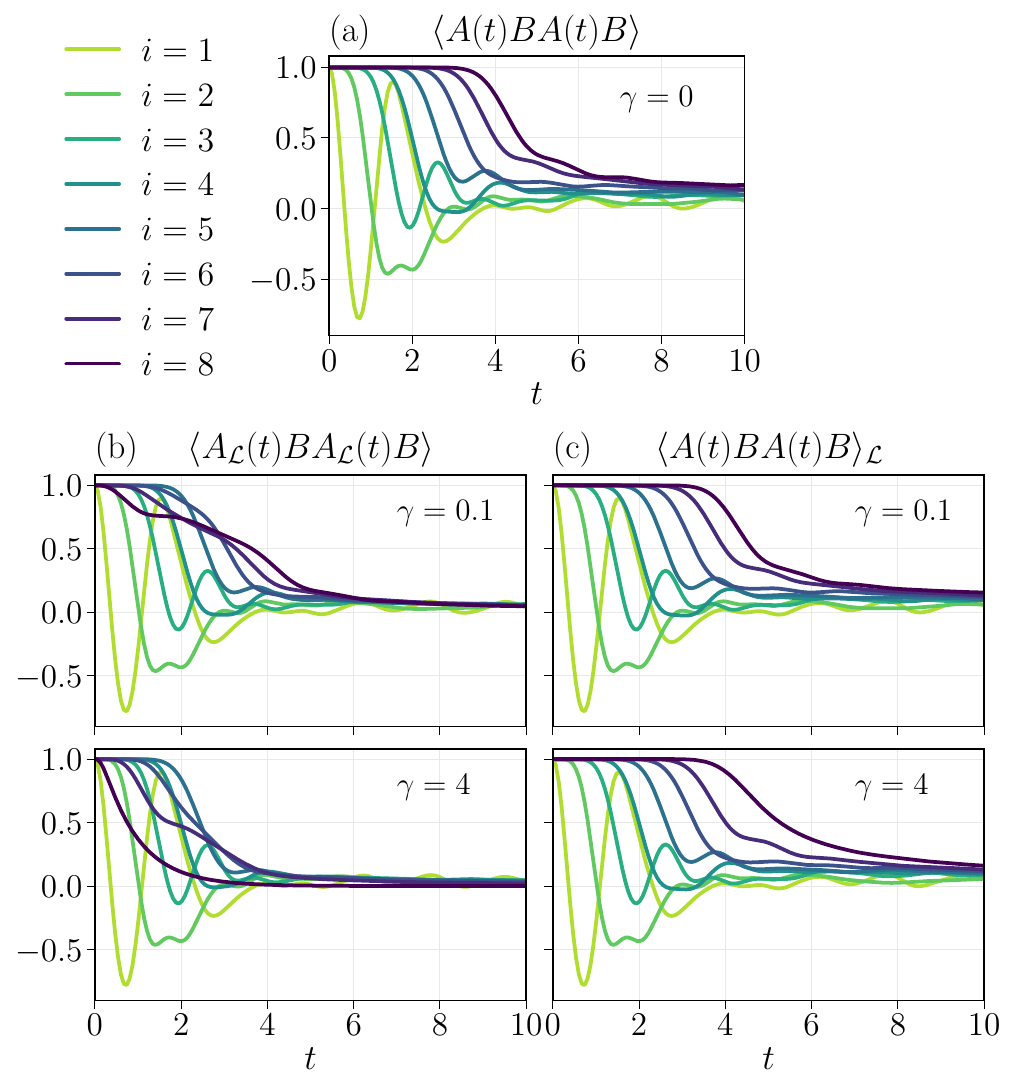}
\caption{Operator spreading in the boundary-dissipated Ising chain~\eqref{chain}.
The OTOCs are numerically studied for $A=\sigma^z_i$, with
$i=1,\ldots,L$, and $B=\sigma^z_1$, for a system of size $L=8$.
Panel (a) shows the OTOC under unitary dynamics, while panels (b) and (c) show the reduced and full OTOCs, respectively, for two different values of $\gamma\neq0$.
The competition between unitary scrambling and dissipation manifests differently in the two quantities, with the full OTOC $\langle A(t)BA(t)B\rangle_{\mathcal L}$ providing a more faithful description of the scrambling dynamics of the corresponding closed system.}
\label{Scrambling}
\end{figure}

\emph{Operator spreading.}
The behavior can also be understood in terms of Pauli strings and operator weight. We denote a Pauli string by $ P_\mu = \bigotimes_{i=1}^L \sigma^i_{\mu_i}$ with $ \mu_i = 0,x,y,z$ and $\sigma_0 \equiv \mathbb{1}$. 
For a qubit chain, an operator can always be expanded in the complete basis of Pauli strings as
$A(t) = \sum_{\mu} a_\mu(t)P_\mu$.
The standard OTOC, for $B$ a single Pauli string, can be written as 
\begin{equation}
    \langle A(t)BA(t)B\rangle = \!\!\!\!\!\! \sum_{\mu:\, [P_\mu,B]=0} \!\!\! |a_\mu(t)|^2 \,\,- \!\!\!\!\!\!\sum_{\mu:\, \{P_\mu,B\}=0} \!\!\!|a_\mu(t)|^2 \ .
\label{OTOC_pauli}
\end{equation}
The OTOC hence measures the relative weight of Pauli strings of $A(t)$ that commute or anticommute with $B$~\cite{xu2024scrambling}.
In the expansion of $A_{\L}(t) = \sum_{\mu} \tilde{a}_\mu(t)P_\mu$, the norm of the operator is not preserved and one finds
\begin{align}
   \frac{d}{dt} \langle A_{\L}(t)A_{\L}(t)\rangle=\frac{d}{dt} \sum_\mu {|\tilde{a}_\mu(t)|^2} =   - 4\gamma \!\!\!\!\sum_{\mu:\, \{P_\mu,J\}=0} \!\!\!\!|\tilde{a}_\mu(t)|^2 \ .
\end{align}
Unlike unitary evolution, dissipative dynamics does not merely redistribute weight among Pauli strings, but also acts as a sink for strings that anticommute with the jump operator. 
Consequently, the reduced OTOC can decay even before the evolved operator reaches the support of $B$, since dissipation removes part of its Pauli-string weight during spreading. Its decay therefore reflects not only scrambling, but also dissipative norm loss, which can mimic signatures of operator spreading.
In this picture, Eq.~\eqref{eq:decay_A2_vs_I} can be rewritten as
$\frac{d}{dt}I_{A,B=\mathbb{1}}(t)=-\frac{d}{dt}\sum_\mu|\tilde a_\mu(t)|^2$.
The replica-correlation term compensates for this loss, restoring the total Pauli-string weight and hence the Hilbert--Schmidt norm. The full OTOC therefore predominantly reflects operator spreading.

\emph{Late-time freeness.} ---
Let us now turn to the long-time limit and study the decay and plateau values of correlations.
OTOCs and their higher-order generalizations have recently been reinterpreted as probing the emergence of free independence in chaotic quantum many-body dynamics, where free independence provides the noncommutative analogue of statistical independence~\cite{cipolloni2022thermalisation,fava2025designs}.
When $A(t)$ becomes freely independent from $B$, as expected under generic dynamics, their asymptotic values are predicted by free probability theory~\cite{voiculescu1992free,mingo2017free,xia2019simple}.
The corresponding predictions are~\footnote{These predictions assume that the observables have no overlap with nontrivial conserved quantities. In the presence of conservation laws, components within the conserved subspace are not fully scrambled and can contribute to the late-time correlations and slow the approach to the asymptotic plateau. The predictions should therefore be understood as applying to the fully scrambled sector, after accounting for possible conserved components.}
\begin{align}
\label{freeness_chain}
&\lim_{t\rightarrow\infty}\braket{A(t)B}
=\langle A\rangle\langle B\rangle \ ,\\
&\lim_{t\rightarrow\infty}\braket{A(t)BA(t)B}
=\langle A^2\rangle\langle B\rangle^2
+\langle B^2\rangle\langle A\rangle^2
-\langle A\rangle^2\langle B\rangle^2 \ . \nonumber
\end{align}
The full OTOC obtained within the extended Lindblad formalism reproduces these predictions analytically~\cite{SuppMat2026}. Indeed, taking the long-time limit of Eqs.~\eqref{eigenop_1} and \eqref{eq:OTOC_full_main} yields Eq.~\eqref{freeness_chain}, since $\lim_{t\rightarrow \infty} A_\mathcal{L}(t) = \langle A \rangle \mathbb{1}$.
The reduced OTOC, instead, relaxes to a different value,
\begin{align}
    \lim_{t \rightarrow\infty} \braket{A_{\L}(t)BA_{\L}(t)B} &= \langle B^2\rangle\langle A\rangle^2 - \langle A\rangle^2\langle B\rangle^2 \nonumber\\
    &+\lim_{t \rightarrow\infty}\langle A_\mathcal{L}(t)A_\mathcal{L}(t)\rangle\langle B\rangle^2 \ ,
\label{freeness_chain_reduced}
\end{align}
and hence decays exponentially toward $\langle B^2\rangle\langle A\rangle^2$.

In Fig.~\ref{Freeness_fig}, we compare the late-time behavior of correlation functions and of the full and reduced OTOCs for systems of different sizes, with and without dissipation. We consider traceful observables $A = B = \tfrac{1}{2}(\mathbb{1}+\sigma_{L/2}^z)$. The correlation functions relax to the expected value $\langle A\rangle^2=1/4$ under both unitary and Lindblad dynamics, with small finite-size deviations in the former that decrease with increasing system size.
Similarly, both the full OTOC and the unitary OTOC approach the prediction of Eq.~\eqref{freeness_chain}, which for these observables gives $2\langle A^2\rangle\langle A\rangle^2-\langle A\rangle^4=3/16$, again up to finite-size corrections in the unitary case. By contrast, the reduced OTOC relaxes to $\langle A^2\rangle\langle A\rangle^2=1/8$, coherently with Eq.~\eqref{freeness_chain_reduced}. Thus, only the full OTOC reproduces the closed-system late-time plateau.
 
\begin{figure}[t]
\centering
\includegraphics[width=\linewidth]{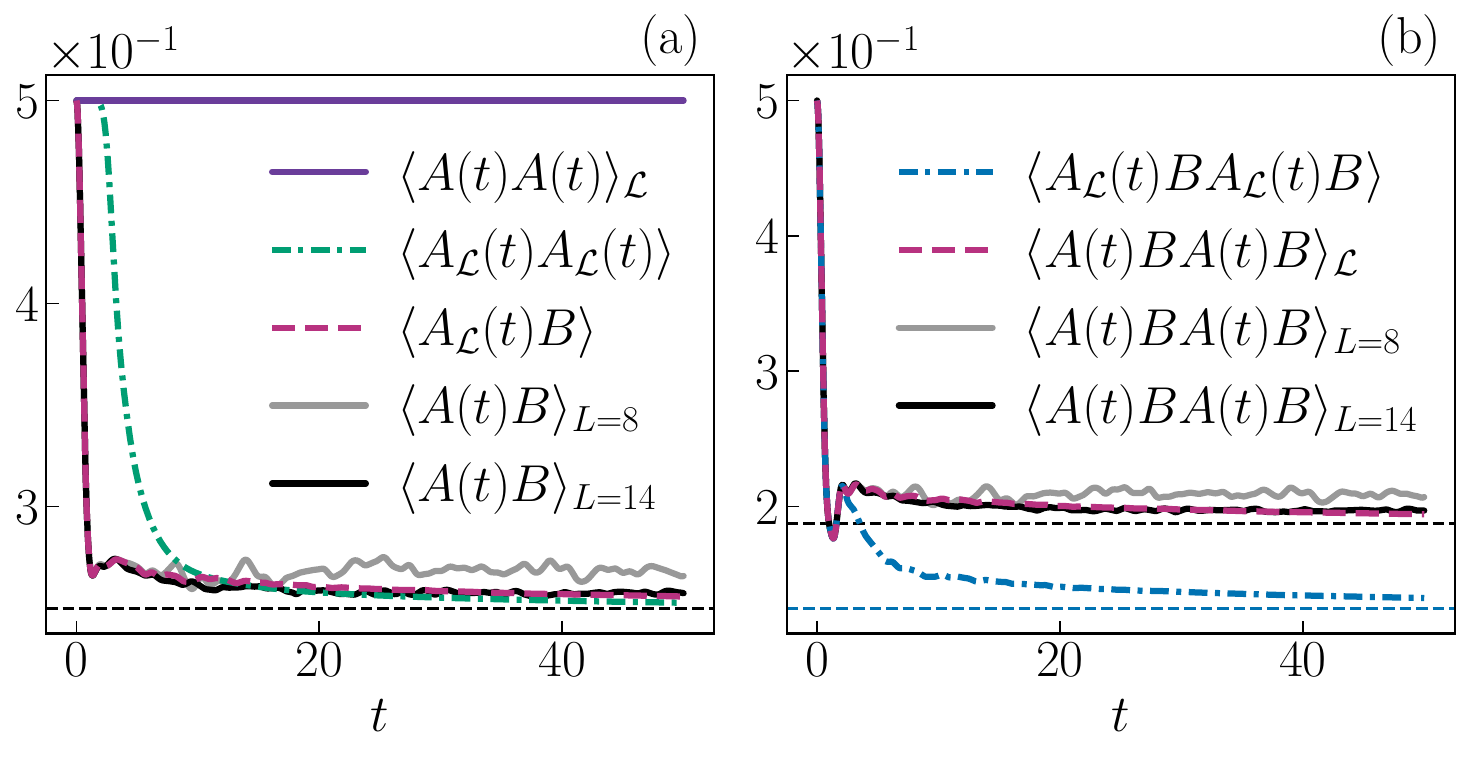}
\caption{
Late-time behavior of correlations between $A=B=\tfrac{1}{2}(\mathbb{1}+\sigma^z_{L/2})$ in the boundary-dissipated Ising chain~\eqref{chain}.
(a) Correlation function for the open system ($\gamma=1$, $L=8$, dashed magenta), compared with the closed system for $L=14$ (solid black) and $L=8$ (solid gray). All curves approach the same plateau $1/4$ (black dashed line).
We also show $\langle A_\mathcal{L}(t)A_\mathcal{L}(t)\rangle$ for $\gamma=1$ (dash-dotted green), clearly distinct from the curve $\langle A(t)A(t)\rangle_\mathcal{L}=\langle A^2\rangle=1/2$ (solid violet). \\
(b) Full OTOC for the open system ($\gamma=1$, $L=8$, dashed magenta), compared with the closed-system OTOCs for $L=14$ (solid black) and $L=8$ (solid gray). These curves approach the same plateau $3/16$ (black dashed line), whereas the reduced OTOC for the open system ($\gamma=1$, $L=8$, dash-dotted blue) approaches $1/8$ (blue dashed line).
The plateau values agree with Eqs.~\eqref{freeness_chain} and~\eqref{freeness_chain_reduced}.}
\label{Freeness_fig}
\end{figure}

\emph{Application.} ---
From the previous discussion, it is clear that the full OTOC retains the spatiotemporal structure of operator spreading observed in unitary dynamics: both the short- and long-time dynamics are accurately reproduced, while finite-size effects are suppressed. By contrast, the reduced OTOC can differ qualitatively from the closed-system dynamics due to dissipation and the associated loss of quantum information.
This is ultimately understood from the fact that $\langle A(t)BA(t)B\rangle_{\L}$ as defined in Eq.~\eqref{C2_full} is constructed from the full unitary description underlying the effective open-system dynamics.
This observation suggests a broader application of our formalism: OTOCs of a large closed system can be approximated by the OTOCs obtained within our formalism for a smaller open system, where the degrees of freedom far from the operators of interest are replaced by a dissipative boundary. 
The connection between open-system dynamics and closed-system correlation functions has already been successfully exploited in the study of quantum transport~\cite{heitmann2023spin,kraft2024lindblad}. The physical intuition is that, in a large closed system, operator weight spreads away from the region of interest into increasingly distant degrees of freedom. In a smaller open system, a dissipative boundary can mimic this process by absorbing the outgoing operator weight, thereby suppressing finite-size reflections and recurrences. Our results extend this idea to OTOCs, provided that the replica-correlation term is included.

As a proof of concept, we consider closed-system dynamics in a `large' system, and compare this with the extended Lindblad dynamics for a `small' system. We consider operator dynamics with $A = B = \sigma_1^z$ located at the left boundary, and in the small system we add dissipation at the right boundary with $J = \sqrt{\gamma} \sigma_L^z$. The approximation is indeed particularly effective for observables located farthest from the dissipative boundary. The value of $\gamma$ is chosen numerically such that the smaller open system optimally approximates the larger system dynamics. Since this value is primarily controlled by the local dynamics near the boundary, it is expected to depend only weakly on system size and can therefore be calibrated on small systems.
We illustrate this idea in Fig.~\ref{HamDiss}, where we show how both the correlation functions and the OTOC for a closed system with $L=14$ can be accurately approximated by the dissipative correlation function $\braket{A(t)B}_{\L}$ and the OTOC $\braket{A(t)BA(t)B}_{\L}$ with $L=8$ and $\gamma = 1$. 
For comparison, we also show the results for the closed system with $L=8$, corresponding to the same system size as the open system.
This result highlights that the inclusion of dissipation allows the full OTOC of a smaller open system to closely reproduce the closed-system OTOC at larger system sizes. The dissipative boundary provides a numerically efficient route to probing operator spreading and information scrambling beyond directly accessible system sizes.

\begin{figure}[t]
\centering
\includegraphics[width=\linewidth]{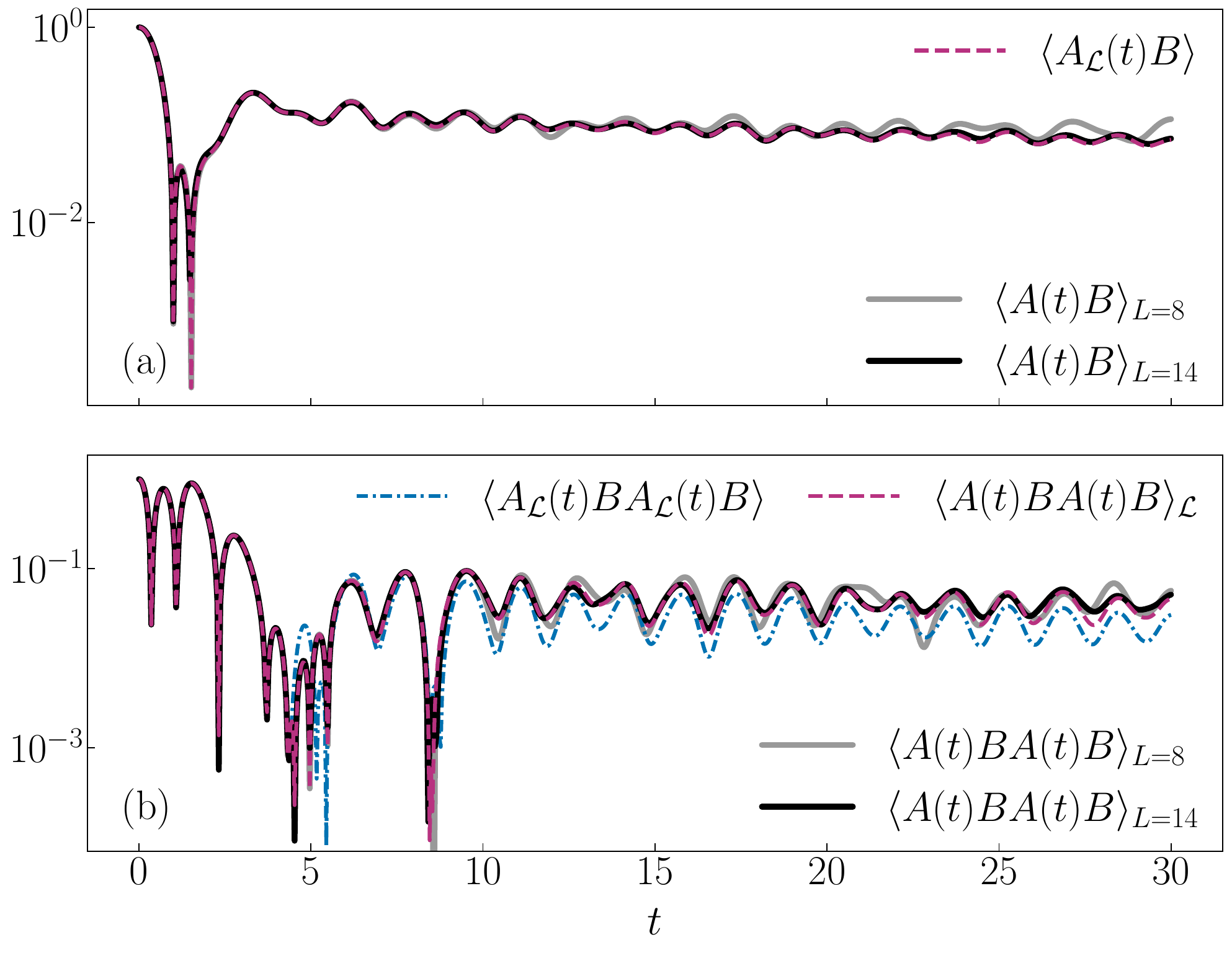}
\caption{
Correlation functions and OTOCs in the boundary-dissipated Ising chain~\eqref{chain}, for boundary operators $A=B=\sigma_1^z$.
(a) Correlation function for the open system ($L=8$, $\gamma=1$, dashed magenta), compared with the ones of the closed system for $L=14$ (solid black) and $L=8$ (solid gray).
(b) Full OTOC for the open system ($L=8$, $\gamma=1$, dashed magenta), compared with the corresponding closed-system OTOCs for $L=14$ (solid black) and $L=8$ (solid gray). The reduced OTOC of the open system is also shown (dash-dotted blue).
In both panels, the open-system results closely reproduce those of the larger closed system, whereas the reduced OTOC provides a significantly poorer approximation.
}
\label{HamDiss}
\end{figure}

\emph{Conclusions and Outlook.} ---
In this work, we introduced a general formalism to study out-of-time-ordered correlators in open quantum systems described by Lindblad dynamics. 
Starting from a microscopic random-circuit model, we have demonstrated that the joint unitary evolution of system and bath introduces corrections to the naive Lindbladian OTOC, stemming from correlations between the two replicas.
The resulting replica-correlation term plays a fundamental role in preserving properties of the underlying unitary dynamics, including conservation of quantum information, such that the full OTOC remains much closer to that of the corresponding closed system, with its intermediate-time decay reflecting operator spreading rather than dissipative loss of operator weight and its late-time behavior recovering the correct plateau associated with free independence.
Beyond these conceptual results, our formalism also suggests a practical computational strategy: OTOCs of larger closed systems can be approximated using smaller systems with suitably engineered dissipative boundaries, providing an efficient route to probing operator growth beyond directly accessible system sizes.

Several possible extensions remain to be explored. 
A first extension concerns higher-order OTOCs, which provide increasingly refined probes of quantum many-body dynamics and have been connected to quantum designs~\cite{dankert2009exact,roberts2017chaos,cotler2017chaos}, deep thermalization~\cite{bhore2023deep,claeys2022emergent,cotler2023emergent}, free independence~\cite{chen2025free,vallini2026longtime,dowling2025free}, and higher-order hydrodynamics~\cite{delacretaz2024nonlinear,doyon2020fluctuations,fava2021hydrodynamic}. Within our framework, averaging over the environment is expected to generate increasingly complex correlations between replicas, giving rise to a hierarchy of additional terms with the structure of free probability naturally built in~\cite{fritzsch2026free}. Understanding this hierarchy could provide a general framework for higher-order dynamical correlations in open quantum systems and help clarify the notion of statistical independence that emerges in the presence of dissipation. 
A second direction concerns the relation between the present formalism and the Eigenstate Thermalization Hypothesis (ETH)~\cite{ srednicki1999approach, dalessio2016from, foini2019eigenstate}. Since our construction is derived from an underlying unitary dynamics, it would be interesting to understand how the replica correlations emerge within an ETH description of the enlarged system, following, e.g., the approaches of Refs.~\cite{odonovan2025master, fritzsch2026freecumulants}. 
Another natural direction is the extension beyond Markovian dynamics. Generalizing our repeated-interaction construction to environments with memory could reveal how non-Markovian effects modify the replica correlations identified here.
Finally, our results point to a broader conceptual message: higher-order correlation functions cannot, in general, be reconstructed from independent copies of Lindblad dynamics. This perspective may extend beyond OTOCs to nonlinear response functions and other probes of quantum information dynamics in open many-body systems. \\

\emph{Acknowledgments.} --- The authors are grateful to Silvia Pappalardi for insightful discussions and helpful suggestions throughout the development of this work.
E.V. thanks Neil Dowling, Beatrice Magni, Philipp Strasberg and all authors thank Gabriel O. Alves for stimulating conversations. 
E.V. is supported by the Deutsche Forschungsgemeinschaft (DFG, German Research Foundation) under Germany’s Excellence Strategy - Cluster of Excellence Matter and Light for Quantum Computing (ML4Q) EXC 2004/1 -390534769, and DFG Collaborative Research Center (CRC) 183 Project No. 277101999 - project B02.
P.W.C. and F.F. acknowledge support from the Max Planck Society.
F.F. further acknowledges support from the European Union's Horizon Europe program under the Marie Sk{\l}odowska Curie Action GETQuantum (Grant No. 101146632).
AI (OpenAI ChatGPT, GPT-5.6 Sol) was used during the preparation of this manuscript to improve clarity and presentation and to assist with code development and optimization. AI was not used to generate the scientific content.

\emph{Data availability.} --- The data that support the findings of this article are openly available~\cite{zenodo2026}.

\bibliographystyle{apsrev4-2}
\bibliography{ref}


\clearpage
\onecolumngrid
\makeatletter
\@booleanfalse\twocolumn@sw
\def\title@column#1{\minipagefootnote@init #1\minipagefootnote@foot}
\def\close@column{\newpage}
\makeatother

\setcounter{secnumdepth}{3}
\setcounter{tocdepth}{2}
\setcounter{section}{0}
\setcounter{subsection}{0}
\setcounter{subsubsection}{0}
\setcounter{figure}{0}
\setcounter{table}{0}
\setcounter{footnote}{0}
\setcounter{equation}{0}
\renewcommand{\theequation}{S\arabic{equation}}

\makeatletter
\frontmatter@init
\let\author\supp@author
\let\affiliation\supp@affiliation
\let\email\supp@email
\let\maketitle\supp@maketitle
\let\thanks\supp@thanks
\let\frontmatter@footnote@produce\frontmatter@footnote@produce@footnote
\makeatother

\title{Supplemental Material for
``Out-of-Time-Ordered Correlators Beyond Lindblad''
}

\author{Elisa Vallini\orcidlink{0009-0003-4613-9526}}
\email{evallini@uni-koeln.de}
\affiliation{Institut f\"ur Theoretische Physik, Universit\"at zu K\"oln, Z\"ulpicher Straße 77, 50937 K\"oln, Germany}

\author{Felix Fritzsch\orcidlink{0000-0002-7659-7574}}
\affiliation{Max Planck Institute for the Physics of Complex Systems, Nöthnitzer Strasse 38, Dresden 01187, Germany}

\author{Pieter W. Claeys\orcidlink{0000-0001-7150-8459}}
\affiliation{Max Planck Institute for the Physics of Complex Systems, Nöthnitzer Strasse 38, Dresden 01187, Germany}
\affiliation{School of Physics, Trinity College Dublin, Dublin 2, Ireland}

\date{\today}

\makeatletter
\let\supp@origlabel\label
\def\label#1{%
  \def\supp@tmp{#1}%
  \def\supp@firstpage{FirstPage}%
  \ifx\supp@tmp\supp@firstpage
    \supp@origlabel{SuppFirstPage}%
  \else
    \supp@origlabel{#1}%
  \fi
}
\makeatother
\maketitle
\makeatletter\let\label\supp@origlabel\makeatother

In this Supplemental Material, we provide additional analysis and technical details supporting the results presented in the main text. In Sec.~\ref{supp_sec_derivation}, we review the minimal circuit model, derive its continuous-time limit, and provide a detailed derivation of the main result of the paper, namely the expression for the ``full'' OTOC including the replica-correlation contribution. We illustrate the formalism in the analytically tractable case of a single qubit in Sec.~\ref{supp_sec_onequbit}. In Sec.~\ref{sec:proof} we provide two additional proofs: we show how the replica correction restores the multiplicative structure of the underlying unitary dynamics, which is lost in the reduced Lindblad description, and we analytically derive late-time freeness for the full OTOC. 

\makeatletter
\newcommand{\supplementtableofcontents}{%
  \begingroup
    \section*{\tocname}%
    \let\appendix\appendix@toc
    \@starttoc{stoc}%
  \endgroup
}

\let\supp@origaddcontentsline\addcontentsline
\newcommand{\supp@redirecttoc}{%
  \def\addcontentsline##1##2##3{%
    \def\supp@ext{##1}%
    \def\supp@tocext{toc}%
    \ifx\supp@ext\supp@tocext
      \supp@origaddcontentsline{stoc}{##2}{##3}%
    \else
      \supp@origaddcontentsline{##1}{##2}{##3}%
    \fi
  }%
}

\supplementtableofcontents
\supp@redirecttoc

\def\@subsectioncntformat#1{\csname the#1\endcsname.\quad}%
\makeatother

\section{Continuous-time dynamics and correlations from the minimal circuit}
\label{supp_sec_derivation}
The starting point of our construction is the minimal circuit model studied in Refs.~[\refippolitiminimal,\reffritzschfree] and summarized in the main text. The model consists of three sites, associated with Hilbert spaces $\mathcal{H}_A$, $\mathcal{H}_C$, and $\mathcal{H}_E$ of dimensions $d_A$, $d_C$ and $d_E$ respectively. The left site $A$ represents the system of interest, which can have arbitrary internal structure such as a spatially extended many-body system, while the right site $E$ represents its environment and acts as an effective bath. The intermediate site $C$ mediates the local coupling between the system and the environment. At each discrete time step, $A$ and $C$ interact through a fixed unitary gate $U_{AC}=U$, while $C$ and $E$ interact through an independent Haar-random unitary $V_{CE}=V_t$. A single time step is therefore described by $\mathcal{U}_t=(\mathbb{1}_A\otimes V_t)(U\otimes\mathbb{1}_E)$, and the evolution up to time $t$ is
\begin{align}
    \mathcal{U}(t)
    = \mathcal{U}_t \dots \mathcal{U}_2\,\mathcal{U}_1
    = \figeq[0.13\columnwidth]{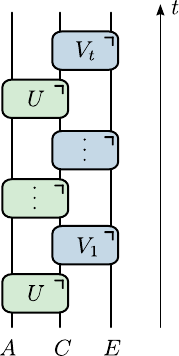} \ .
\end{align}
In the limit $d_E \gg d_A , d_C$, the interaction with the random bath effectively resets the site $C$ after each time step. As a consequence, at each step of the dynamics, the system $A$ behaves as if it were interacting with a freshly prepared ancilla. In this sense, the model can be viewed as a repeated-interaction model, in which the system $A$ interacts sequentially with effectively independent realizations of $C$, each acting as a fresh effective bath. Correlations generated between $A$ and $C$ during one interaction therefore do not persist to subsequent time steps, resulting in a Markovian reduced dynamics for $A$. This Markovian structure allows for a well-defined continuous-time limit, in which the reduced dynamics is described by a quantum master equation of Lindblad form. In this way, the repeated-interaction structure provides a natural microscopic realization of the resulting Lindblad dynamics.

\subsection{Lindblad dynamics}
We next show explicitly how the Markovian reduced dynamics emerges for observables supported on $A$ and how the Lindblad evolution is recovered in the continuous-time limit.

We consider an observable $\mathcal{A}$ of the full circuit, supported only on the system $A$, i.e., $\mathcal{A}=A\otimes\mathbb{1}_C\otimes\mathbb{1}_E$, where $A$ acts on the system Hilbert space of dimension $D\equiv d_A$ (in the main text we used the notation $A_A\equiv \mathcal{A})$. Throughout this work, we use the convention $\mathcal{U}\mathcal{A}\,\mathcal{U}^\dagger$ for the Heisenberg evolution on the full circuit. As shown in Ref.~[\reffritzschfree], after averaging over the random environment and taking the limit $d_E\rightarrow\infty$, the evolution of $\mathcal{A}$ can be described entirely in terms of a reduced map $\mathcal{M}$ acting on $A$ during each discrete time step as
\begin{equation}\label{supp:ev_1op}
    \mathcal{M}(A)=\frac{1}{d_C}\operatorname{Tr}_C\!\left[U(A\otimes\mathbb{1}_C)U^\dagger\right]  .
\end{equation}
The map $\mathcal{M}$ is completely positive and unital and describes the averaged reduced Heisenberg evolution of observables supported on the system $A$. We take Eq.~\eqref{supp:ev_1op} as the starting point for deriving the continuous-time reduced dynamics. 
To obtain the continuous-time limit, we associate a duration $\Delta t$ with each discrete time step and parametrize the corresponding $A$--$C$ interaction as
\begin{align}
    U_{\Delta t}
    =\exp\left[
    -i\Delta t\left(
    H_A\otimes\mathbb{1}_C
    +\mathbb{1}_A\otimes H_C
    \right)
    -i\sqrt{\Delta t}
    \sum_{\alpha}J^{(\alpha)}_{A}\otimes J^{(\alpha)}_{C}
    \right]  ,
\label{eq:unitary_gate_continuous}
\end{align}
where $H_A$ and $H_C$ are traceless Hermitian operators acting on $A$ and $C$, respectively, and $J^{(\alpha)}_{A}$ and $J^{(\alpha)}_{C}$ are traceless Hermitian operators. For convenience, we choose the operators acting on $C$ to be orthonormal,
\begin{equation}\label{orthonormal}
    \frac{1}{d_C}\operatorname{Tr}_C\left[J^{(\alpha)}_{C}J^{(\beta)}_{C}\right]=\delta_{\alpha \beta}\ .
\end{equation}
This choice fixes the normalization of the corresponding jump operators and can be relaxed by absorbing the resulting correlation matrix into their definition.\\
The reduced channel associated with $U_{\Delta t}$ will be denoted by $\mathcal{M}_{\Delta t}$.
Introducing
\begin{equation}
    H_{0}=H_A\otimes\mathbb{1}_C+\mathbb{1}_A\otimes H_C \ , \qquad
    W_0=\sum_{\alpha}J^{(\alpha)}_{A}\otimes J^{(\alpha)}_{C}\ , \qquad \mathbb{1} = \mathbb{1}_A \otimes \mathbb{1}_C\ ,
\end{equation}
the gate can be expanded for small $\Delta t$ as
\begin{equation}
    U_{\Delta t}
    =\mathbb{1}
    -i\sqrt{\Delta t}\,W_0
    -i\Delta t H_0
    -\frac{\Delta t}{2}W_0^2
    +\mathcal{O}(\Delta t^{3/2}) \ .
\end{equation}
Substituting this expansion into Eq.~\eqref{supp:ev_1op} gives the reduced evolution over a single time step,
\begin{align}
    \mathcal{M}_{\Delta t}(A)
    ={}&A
    -i\sqrt{\Delta t}\,
    \frac{1}{d_C}\operatorname{Tr}_C \left(
    \left[W_0,A\otimes\mathbb{1}_C\right]
    \right)
    \nonumber\\
    &+\Delta t\,
    \frac{1}{d_C}\operatorname{Tr}_C
    \left(
    -i[H_0,A\otimes\mathbb{1}_C]
    +W_0(A\otimes\mathbb{1}_C)W_0
    -\frac{1}{2}\left\{
    W_0^2,A\otimes\mathbb{1}_C
    \right\}
    \right)
    +\mathcal{O}(\Delta t^{3/2}) \ .
\end{align}
The term of order $\sqrt{\Delta t}$ vanishes because the operators $J^{(\alpha)}_C$ are traceless. Using their orthonormality, Eq.~\eqref{orthonormal}, and keeping only terms up to order $\Delta t$ yields
\begin{equation}
    \mathcal{M}_{\Delta t}(A)
    =A+\Delta t\,\mathcal{L}(A)
    +\mathcal{O}(\Delta t^{3/2}) \ ,
\end{equation}
where
\begin{equation}
    \mathcal{L}(A)
    =-i[H,A]
    +\sum_{\alpha}
    \left(
    J_{\alpha}AJ_{\alpha}
    -\frac{1}{2}\{J_{\alpha}^2,A\}
    \right)\ ,
    \qquad
    J_{\alpha}\equiv J^{(\alpha)}_{A}, \qquad H\equiv H_A\ .
\end{equation}
After $n$ time steps, we take the continuous-time limit $\Delta t\rightarrow0$ and $n\rightarrow\infty$ while keeping the physical time $t=n\Delta t$ fixed. Using the expansion above and the standard
exponential limit, we obtain
\begin{equation}\label{continuous_limit}
     \mathcal{M}_{\Delta t}^{\,n}(A)\longrightarrow  e^{t\mathcal{L}}(A) =: A_{\mathcal{L}}(t)\ .
\end{equation}
Equivalently, the observable evolved through the repeated application of the channel $\mathcal{M}$ satisfies the Lindblad equation $\partial_t A_{\mathcal{L}}=\mathcal{L}\!\left(A_{\mathcal{L}}\right)$.

It is important to stress that the operators $J_{\alpha}$ can be chosen arbitrarily. If all of them vanish, the system $A$ evolves unitarily, as it is decoupled from the site $C$. Conversely, when the interaction is non-zero, the coupling between $A$ and $C$, combined with the averaging over the random bath $E$, gives rise to dissipative terms in the reduced dynamics of $A$, with the operators $J_{\alpha}$ becoming the corresponding jump operators in the continuous-time limit. This highlights the role of $E$ as an effective Markovian bath, a property that naturally emerges from the repeated-interactions picture reproduced in the simple circuit model.

\subsection{Correlation functions}

At the level of two-point correlation functions, the averaged dynamics of the full circuit is completely determined by the reduced single-system channel $\mathcal{M}_{\Delta t}$.
Having established the continuous-time reduced dynamics of a single observable under the repeated application of this channel, we can now evaluate the correlation function of the full circuit between two observables $\mathcal{A}=A\otimes\mathbb{1}_C\otimes\mathbb{1}_E$ and $\mathcal{B}=B\otimes\mathbb{1}_C\otimes\mathbb{1}_E$ supported on the system. 
After $n$ discrete time steps and in the thermodynamic limit, the bath-averaged correlation function is given by
\begin{equation}
    C^{(1)}_{AB}(n)
    =
    \lim_{d_E\rightarrow\infty}
    \mathbb{E}\!\left[
        \left\langle\mathcal{A}(n)\mathcal{B}\right\rangle
    \right]
    =
    \left\langle\mathcal{M}_{\Delta t}^{\,n}(A)B\right\rangle \ .
\end{equation}
Here, the average $\mathbb{E}$ is taken with respect to the $n$ independent Haar-random unitaries $V_i$ and $d_E \to \infty$ models the thermodynamic limit for the environment.
It is then immediate to take the continuous-time limit. Using Eq.~\eqref{continuous_limit},
\begin{equation}
    C^{(1)}_{AB}(n) \longrightarrow
    \left\langle A_{\mathcal{L}}(t)B\right\rangle, \qquad A_{\mathcal{L}}(t)=e^{t\mathcal{L}}(A)  \ .
\end{equation}
Hence, the correlation function of the full unitary circuit in the thermodynamic and continuous-time limit is completely determined by the reduced Lindblad dynamics on $A$.
For later use, we also introduce the adjoint evolution
\begin{equation}
    B_{\mathcal{L}^\dagger}(t)= e^{t\mathcal{L}^{\dagger}}(B) \ ,
\end{equation}
defined through
\begin{equation}
    \left\langle e^{t\mathcal{L}}(A)B\right\rangle
    =
    \left\langle A e^{t\mathcal{L}^{\dagger}}(B)\right\rangle
    =
    \left\langle AB_{\mathcal{L}^\dagger}(t)\right\rangle \ .
\end{equation}

\subsection{Two-replica dynamics and derivation of the OTOC}
\label{supp_sec_OTOC}

We now derive the continuous-time expression for the OTOC starting from the two-replica dynamics of the minimal circuit model, recalling only the ingredients of the discrete-time construction that are needed below and referring to Ref.~[\reffritzschfree] for the general formalism.
Considering again two observables $\mathcal{A}=A\otimes\mathbb{1}_C\otimes\mathbb{1}_E$ and $\mathcal{B}=B\otimes\mathbb{1}_C\otimes\mathbb{1}_E$ supported on the system, we define the bath-averaged OTOC after $n$ discrete time steps as
\begin{equation}
    C_{AB}^{(2)}(n)
    =
    \lim_{d_E\rightarrow\infty}
    \mathbb{E}\!\left[
        \left\langle
        \mathcal{A}(n)\mathcal{B}\mathcal{A}(n)\mathcal{B}
        \right\rangle
    \right] \ .
    \label{eq:supp_otoc_discrete}
\end{equation}
While the two-point correlation function is completely determined by repeated applications of the single-copy reduced channel $\mathcal{M}$, the OTOC contains two copies of the time-evolved observable. It is therefore useful to introduce two replicas of the circuit and describe their joint evolution.

The corresponding replica picture is obtained by folding the copies of the evolution operator and its inverse on top of each other in an alternating fashion. We introduce the two-folded gate
\begin{equation}
    \figeq[0.06\columnwidth]{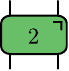}
    \,\,=\frac{1}{d_C^2}\,\,
    \figeq[0.1\columnwidth]{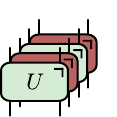}\ ,
    \qquad 
    U^{(2)}
    =
    \frac{1}{d_C^2}
    \left(U\otimes U^*\right)^{\otimes 2} \ ,
    \qquad
    U^{(2)}:
    \mathcal{H}_A^{\otimes 4}\otimes\mathcal{H}_C^{\otimes 4}
    \longrightarrow
    \mathcal{H}_A^{\otimes 4}\otimes\mathcal{H}_C^{\otimes 4}\ .
\end{equation}
It acts on two replicas, each consisting of a forward and a backward time sheet. Here, $*$ denotes complex conjugation, and the green and red gates denote $U$ and $U^*$, respectively.
It is useful to introduce permutation states on the replicated forward--backward space of a subsystem $X$, where $X$ can be either $A$ or $C$. We denote these states by $|\sigma)_X$, for a permutation $\sigma\in S_2$. For two replicas, the replicated forward--backward space is $\mathcal{H}_X^{\otimes 4}$, and there are two permutation states,
\begin{equation}
    |\circ)_X = \sum_{i,j=1}^{d_X} \ket{iijj}_X\ ,\qquad |\ssquare)_X = \sum_{i,j=1}^{d_X} \ket{ijji}_X\ ,
\end{equation}
with $\ket{i}_X, i=1 \dots d_X$ an orthonormal basis for the Hilbert space $\mathcal{H}_X$. These permuation states are associated with the identity permutation $\sigma=\circ=(1)(2)$ and the swap permutation $\sigma=\ssquare=(12)$, respectively. The identity pairs each forward sheet with the backward sheet of the same replica, whereas the swap exchanges the two pairings, connecting the forward sheet of one replica with the backward sheet of the other. These permutation states enter in two distinct places: for $X=C$, they arise when performing the average over the environment, whereas for $X=A$, they enter in the boundary conditions defining the OTOC (we refer to Ref.~[\reffritzschfree] for a more detailed discussion).

As shown in Ref.~[\reffritzschfree], averaging over the random environment introduces an additional label in replica space, which keeps track of how the forward and backward sheets are paired through the environment. More precisely, this label specifies the contraction pattern of the replicated $C$ degrees of freedom generated by the average. In the thermodynamic limit $d_E\to\infty$, the surviving contraction patterns are labeled by non-crossing partitions. For two replicas, there are only two such configurations, corresponding to the two permutations introduced above, $\circ$ and $\ssquare$. More explicitly, the identity configuration connects each forward $C$ sheet with the backward $C$ sheet of the same replica, whereas the swap configuration exchanges these connections, pairing the forward $C$ sheet of one replica with the backward $C$ sheet of the other.
After averaging over the environment, the reduced two-replica dynamics on $A$ is described by two coupled components, corresponding to the two possible contraction patterns $\ssquare$ and $\circ$ of the replicated $C$ degrees of freedom.
The corresponding discrete-time evolution is governed by the triangular transfer matrix
\begin{equation}\label{transfer_matrix}
\mathcal{T}=
\left(
\begin{array}{c|c}
    \mathcal{M}_{\ssquare\ssquare}
    &
    d_C\mathcal{M}_{\ssquare\circ}-\mathcal{M}_{\ssquare\ssquare}
    \\ \hline
    0
    &
    \mathcal{M}_{\circ\circ}
\end{array}
\right)\ , \qquad 
    \mathcal{T}:
    \left(\mathcal{H}_A^{\otimes4}\right)
    \oplus
    \left(\mathcal{H}_A^{\otimes4}\right)
    \longrightarrow
    \left(\mathcal{H}_A^{\otimes4}\right)
    \oplus
    \left(\mathcal{H}_A^{\otimes4}\right) .
\end{equation}
The triangular structure of $\mathcal{T}$ reflects the partial ordering of the non-crossing-partition lattice associated with the replicated $C$ degrees of freedom. Indeed, the environmental average allows the dynamics to proceed only along non-decreasing paths on this lattice. The two permutations are ordered as $\circ <\ssquare$, so that the dynamics can remain within either sector or evolve from $\circ$ to $\ssquare$, but not in the opposite direction. This is precisely encoded by the vanishing lower-left block of $\mathcal{T}$.

Each block $\mathcal{M}_{\nu\sigma}$, with $\nu,\sigma\in\{\circ,\ssquare\}$, is a linear map $\mathcal{M}_{\nu\sigma}: \mathcal{H}_A^{\otimes4} \longrightarrow \mathcal{H}_A^{\otimes4}$, where $\sigma$ and $\nu$ specify respectively the incoming and the outgoing contraction patterns of the replicated $C$ degrees of freedom. Explicitly,
\begin{equation}\label{maps_M}
\mathcal{M}_{\nu\sigma}
    = \,\, \figeq[0.09\columnwidth]{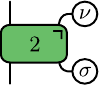} \ , \qquad 
    \mathcal{M}_{\nu\sigma}
    =
    \left(
        \mathbb{1}_A^{\otimes4}\otimes{}_C(\nu|
    \right)
    U^{(2)}
    \left(
        \mathbb{1}_A^{\otimes4}\otimes|\sigma)_C
    \right)\ .
\end{equation}
In this representation, the OTOC is obtained as a matrix element of the transfer matrix $\mathcal{T}$ between an initial and a final boundary vector. The boundary states therefore encode both the insertions of the observables $A$ and $B$, respectively, and the corresponding contraction pattern of the replicated $A$ degrees of freedom. Indeed, after folding the circuit, the operator product entering the OTOC is reproduced by appropriately contracting the open forward and backward $A$ legs at the two ends of the replicated evolution.

We define the initial and final boundary states on $\mathcal{H}_A^{\otimes 4}$ as
\begin{align}
    |\bullet)
    &=
    \left(
        A\otimes\mathbb{1}_A\otimes
        A\otimes\mathbb{1}_A
    \right)
    |\circ) \ ,
    \qquad \qquad   
    (\blackssquare|
    =
    (\ssquare|
    \left(
        \mathbb{1}_A\otimes B\otimes
        \mathbb{1}_A\otimes B
    \right) \ .
\end{align}
Here, the identity contraction $|\circ)$ closes each replica independently at the initial boundary, while the swap contraction $(\ssquare|$ connects the two replicas at the final boundary. This difference in the boundary contractions is what reproduces the trace structure of the OTOC in the folded representation. In particular, the final swap contraction joins the replicated $A$ legs into a single contraction cycle, yielding the ordered product $\langle A(n)BA(n)B\rangle$. An identity contraction at the final boundary would instead close the two replicas separately, resulting in a product of two independent traces.
These boundary contractions act on the replicated $A$ degrees of freedom and should be distinguished from the $\circ$ and $\ssquare$ contraction sectors of the replicated $C$ degrees of freedom that label the components of the transfer matrix.
To contract these boundary states with the transfer matrix, they must also be resolved with respect to the two $C$-replica sectors. The corresponding initial and final vectors are
\begin{equation}
    |\psi_A))
    =
    \begin{pmatrix}
        |\bullet)\\
        |\bullet)
    \end{pmatrix} \ ,
    \qquad \qquad 
    ((\psi_B|
    = \frac{1}{d_A}
    \begin{pmatrix}
        (\blackssquare| & 0 \,\,
    \end{pmatrix} .
\end{equation}
The initial boundary vector has components in both $C$-replica sectors, whereas the final vector has support only on the first ($\ssquare$) sector.
Indeed, as mentioned above, while the dynamics may start from either contraction sector and follows non-decreasing paths on the partition lattice, it is constrained to terminate in the cyclic sector. 
Nevertheless, both components of the initial state must be retained, since they are coupled during the evolution by the off-diagonal block of $\mathcal{T}$. 

The OTOC after $n$ discrete time steps is given by
\begin{equation}
    C_{AB}^{(2)}(n)
    = 
    ((\psi_B|\,\mathcal{T}^{\,n}\,|\psi_A)) \ .
\end{equation}
To obtain the continuous-time dynamics, we associate a duration $\Delta t$
with each discrete time step. Expanding the three maps entering
$\mathcal{T}_{\Delta t}$ to first order in $\Delta t$, as shown below, the
two-replica transfer matrix takes the form
\begin{equation}\label{con_time_2replica}
    \mathcal{T}_{\Delta t}
    =
    \mathbb{1}
    +
    \Delta t\,\mathcal{R}
    +
    \mathcal{O}(\Delta t^{3/2}) \ ,
    \qquad \quad 
    \mathcal{R}
    =
    \left(
    \begin{array}{c|c}
        \mathcal{R}_{\ssquare\ssquare} &
        \mathcal{R}_{\ssquare\circ}
        \\ \hline
        0 &
        \mathcal{R}_{\circ\circ}
    \end{array}
    \right) \ .
\end{equation}
Taking $\Delta t\rightarrow0$ and $n\rightarrow\infty$ while keeping the
physical time $t=n\Delta t$ fixed gives $\mathcal{T}_{\Delta t}^{\,n}  \longrightarrow e^{t\mathcal{R}}$.
Owing to its triangular structure, the corresponding propagator can be written explicitly as
\begin{equation}
    e^{t\mathcal{R}}
    =
    \left(
    \begin{array}{c|c}
        e^{t\mathcal{R}_{\ssquare\ssquare}}
        &
        \displaystyle
        \int_0^t ds\,
        e^{(t-s)\mathcal{R}_{\ssquare\ssquare}}
        \mathcal{R}_{\ssquare\circ}
        e^{s\mathcal{R}_{\circ\circ}}
        \\[10pt]  \hline
        \\[-8pt]
        0
        &
        e^{t\mathcal{R}_{\circ\circ}}
    \end{array}
    \right) \ .
\end{equation}
The continuous-time evolution of the initial boundary vector $|\psi_A))$ is therefore generated by $\mathcal{R}$ and, at the final time, contracted with the final boundary vector $((\psi_B|$. We denote the resulting continuous-time OTOC by $\langle A(t)BA(t)B\rangle_{\mathcal{L}}$:
\begin{align}
   C_{AB}^{(2)}(n) =(( \psi_B|  \mathcal{T}_{\Delta t}^{\,n} |\psi_A )) \  \longrightarrow \ & \langle A(t)BA(t)B\rangle_{\mathcal{L}} = \frac{1}{d_A}
    \begin{pmatrix}
        (\blackssquare|\quad 0\,\,
    \end{pmatrix}
    \left(
    \begin{array}{c|c}
        e^{t\mathcal{R}_{\ssquare\ssquare}}
        &
        \displaystyle
        \int_0^t ds\,
        e^{(t-s)\mathcal{R}_{\ssquare\ssquare}}
        \mathcal{R}_{\ssquare\circ}
        e^{s\mathcal{R}_{\circ\circ}}
        \\[10pt]  \hline
        \\[-8pt]
        0
        &
        e^{t\mathcal{R}_{\circ\circ}}
    \end{array}
    \right)
    \begin{pmatrix}
        |\bullet)\\
        |\bullet)
    \end{pmatrix} \\ 
    &=\frac{1}{d_A} \,
    (\blackssquare| e^{t\mathcal{R}_{\ssquare\ssquare}} |\bullet) + \frac{1}{d_A}
    \int_0^t ds\, (\blackssquare|
        e^{(t-s)\mathcal{R}_{\ssquare\ssquare}}
        \mathcal{R}_{\ssquare\circ}
        e^{s\mathcal{R}_{\circ\circ}}|\bullet) \ .
\label{OTOC_cont}
\end{align}
To evaluate the two contributions in Eq.~\eqref{OTOC_cont}, we first note that the diagonal replica sectors simplify when acting on the corresponding boundary states.

\begin{itemize}
    \item In the $\circ\circ$ sector, the contraction pattern separates the two replicas, so that the evolution factorizes into two independent single-replica contributions, as graphically evident 
\begin{align}
    \figeq[0.08\columnwidth]{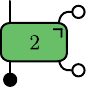}
    \quad = \quad
    \figeq[0.12\columnwidth]{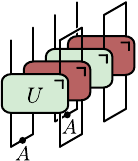}
    \quad = \quad
    \figeq[0.08\columnwidth]{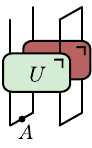}
    \otimes
    \figeq[0.08\columnwidth]{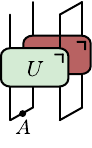}\ .
\end{align}
The resulting evolution is therefore factorized into two independent applications of the single-copy reduced channel,
\begin{equation}
    \mathcal{M}_{\circ\circ}|\bullet)
    =
    \left[
        \mathcal{M}(A)\otimes\mathbb{1}_A
        \otimes
        \mathcal{M}(A)\otimes\mathbb{1}_A
    \right]
    |\circ)_A \ .
\end{equation}
In the continuous-time limit, this gives
\begin{equation}
    e^{t\mathcal{R}_{\circ\circ}}|\bullet)
    =
    \left[
        A_\mathcal{L}(t)\otimes\mathbb{1}_A
        \otimes
        A_\mathcal{L}(t)\otimes\mathbb{1}_A
    \right]
    |\circ)
    \equiv
    |\bullet^{}_t) ,
    \qquad
    A_\mathcal{L}(t)=e^{t\mathcal{L}}(A)\ .
    \label{eq:supp_circle_factorization}
\end{equation}

\item  Similarly, when the $\ssquare\ssquare$ block is contracted from the left with the final boundary state, the contraction pattern again separates the two replicas, leading to two independent single-replica contributions,
\begin{align}
    \figeq[0.08\columnwidth]{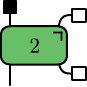}
    \quad = \quad
    \figeq[0.13\columnwidth]{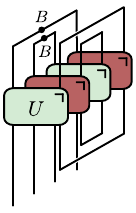}
    \quad = \quad
    \figeq[0.12\columnwidth]{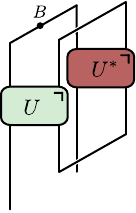}
    \otimes
    \figeq[0.08\columnwidth]{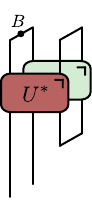}\ .
\end{align}
The resulting evolution is therefore given by two independent adjoint single-copy channels, 
\begin{equation}
    (\blackssquare|
    \mathcal{M}_{\ssquare\ssquare}
    =
    (\ssquare|
    \left[
        \mathbb{1}_A\otimes\mathcal{M}^{\dagger}(B)
        \otimes
        \mathbb{1}_A\otimes\mathcal{M}^{\dagger}(B)
    \right] \ .
\end{equation}
In the continuous-time limit, this gives
\begin{equation}
    (\blackssquare|
    e^{t\mathcal{R}_{\ssquare\ssquare}}
    =
    (\ssquare|
    \left[
        \mathbb{1}_A\otimes B_{\mathcal{L}^\dagger}(t)
        \otimes
        \mathbb{1}_A\otimes B_{\mathcal{L}^\dagger}(t)
    \right]
    \equiv
    (\blackssquare^{}_{\,t}| \ ,
    \qquad
    B_{\mathcal{L}^\dagger}(t)=e^{t\mathcal{L}^{\dagger}}(B)\  .
    \label{eq:supp_square_factorization}
\end{equation}
\end{itemize}
Using these two identities, Eq.~\eqref{OTOC_cont} becomes
\begin{align}
    \langle A(t)BA(t)B\rangle_{\mathcal{L}}
    ={}& \frac{1}{d_A}
    (\blackssquare^{}_{\,t}|\bullet)
    + \frac{1}{d_A}
    \int_0^t ds\,
    (\blackssquare^{}_{\,t-s}|
    \mathcal{R}_{\ssquare\circ}
    |\bullet^{}_{s}) \ .
    \label{OTOC_cont2}
\end{align}

To evaluate the two contributions in Eq.~\eqref{OTOC_cont2}, it is useful to translate the replica notation back into an operator representation on the system Hilbert space $\mathcal{H}_A$. This allows us to express both terms directly in terms of the reduced Lindblad dynamics and the jump operators.

The first term is obtained by contracting the diagonally propagated $\ssquare$ component with the initial boundary state. It therefore corresponds to a contribution in which the replica configuration remains in the $\ssquare$ sector throughout the evolution, that is 
\begin{equation}
    \frac{1}{d_A}
    (\blackssquare^{}_{\,t}|\bullet)
    =
    \left\langle
        A B_{\mathcal L^\dagger}(t) A B_{\mathcal L^\dagger}(t)
    \right\rangle \ , \qquad \qquad B_{\mathcal L^\dagger}(t)=e^{t\mathcal L^\dagger}(B) \ .
    \label{eq:supp_diagonal_otoc_2}
\end{equation}

To evaluate the second term in Eq.~\eqref{OTOC_cont2}, we need the explicit action of the off-diagonal generator $\mathcal{R}_{\ssquare\circ}$. According to Eq.~\eqref{transfer_matrix}, this block originates from the combination $d_C\mathcal{M}_{\ssquare\circ}-\mathcal{M}_{\ssquare\ssquare}$.
We therefore determine its continuous-time limit by expanding the corresponding discrete replica maps to first order in $\Delta t$.
For this purpose, we first rewrite the action of the discrete replica maps $\mathcal{M}_{\nu\sigma}: \mathcal{H}_A^{\otimes4}\longrightarrow\mathcal{H}_A^{\otimes4}$ on the initial boundary state in terms of operators acting on $\mathcal{H}_A^{\otimes2}$. Indeed, this is precisely the action that enters Eq.~\eqref{OTOC_cont2}, where the maps ultimately act on the evolved initial boundary state. For this specific boundary
structure, we denote this operator representation by $ \widehat{\mathcal{M}}_{\nu\sigma}:
\operatorname{End}(\mathcal{H}_A^{\otimes2})
\longrightarrow
\operatorname{End}(\mathcal{H}_A^{\otimes2})$.
Explicitly, one can derive 
\begin{equation}\label{M_unfolded}
    \widehat{\mathcal{M}}_{\nu\sigma}\!\left(A^{\otimes2}\right)
    =
    \frac{1}{d_C^2}
    \operatorname{Tr}_{C^{\otimes2}}
    \left[
        U^{\otimes2}
        \left(
            A^{\otimes2}\otimes P_\sigma^{-1}
        \right)
        U^{\dagger\otimes2}
        \left(
            \mathbb{1}_A^{\otimes2}\otimes P_\nu
        \right)
    \right] \ .
\end{equation}
Here $P_\sigma$ and $P_\nu$ act on the two replicated copies of $C$ by permuting the two tensor factors according to the permutations $\sigma$ and $\nu$, respectively.
We stress that Eq.~\eqref{M_unfolded} represents the action of $\mathcal{M}_{\nu\sigma}$ for the specific initial boundary considered here; a different boundary state would in general lead to a different operator representation. Thus, $\widehat{\mathcal{M}}_{\nu\sigma}$ should not be understood as an additional dynamical map, but only as a convenient notation for this particular action of $\mathcal{M}_{\nu\sigma}$.

Eq.~\eqref{M_unfolded} allows us to insert directly the parametrization of $U_{\Delta t}$ introduced in Eq.~\eqref{eq:unitary_gate_continuous} and expand the replica maps to first order in $\Delta t$. Although only the combination entering
$\mathcal{R}_{\ssquare\circ}$ will ultimately be needed, we report the
three relevant expansions for completeness.

\begin{itemize}

\item We first consider the $\circ\circ$ channel. In this case, the two replicas factorize, as already discussed above, and the operator representation reduces to two independent copies of the single-replica
channel,
\begin{align}
    \widehat{\mathcal{M}}_{\circ\circ}\!\left(A^{\otimes2}\right)
    &=
    \mathcal{M}_{\Delta t}(A)
    \otimes
    \mathcal{M}_{\Delta t}(A)
    =
    A\otimes A
    +\Delta t\left[
        \mathcal{L}(A)\otimes A
        +A\otimes\mathcal{L}(A)
    \right]
    +\mathcal{O}(\Delta t^{3/2}) \ .
\end{align}

\item For the $\ssquare\ssquare$ channel, the corresponding expansion gives
\begin{align}
    \widehat{\mathcal{M}}_{\ssquare\ssquare}\!\left(A^{\otimes2}\right)
    ={}&
    \widehat{\mathcal{M}}_{\circ\circ}
    +\Delta t\sum_{\alpha}
    \Big[
        J_{\alpha}A\otimes AJ_{\alpha}
        +AJ_{\alpha}\otimes J_{\alpha}A
        -J_{\alpha}AJ_{\alpha}\otimes A
        -A\otimes J_{\alpha}AJ_{\alpha}
    \Big]
    +\mathcal{O}(\Delta t^{3/2}) \ .
    \label{eq:supp_Mss}
\end{align}
Notice that, although the action of $\mathcal{M}_{\ssquare\ssquare}$ on the initial boundary does not factorize in the same way as in the $\circ\circ$ sector, its action from the left on the final boundary does factorize, as discussed above.

\item Finally, for the mixed channel we obtain
\begin{equation}
    \widehat{\mathcal{M}}_{\ssquare\circ}\!\left(A^{\otimes2}\right)
    =
    \frac{1}{d_C}\widehat{\mathcal{M}}_{\circ\circ}\!\left(A^{\otimes2}\right)
    -\frac{\Delta t}{d_C}
    \sum_{\alpha}
    [A,J_{\alpha}]\otimes[A,J_{\alpha}]
    +\mathcal{O}(\Delta t^{3/2}) \ .
\end{equation}

\end{itemize}
Therefore, the off-diagonal combination entering the transfer matrix, Eq.~\eqref{transfer_matrix}, has the operator representation
\begin{align}
    d_C\widehat{\mathcal{M}}_{\ssquare\circ}\!\left(A^{\otimes2}\right)
    -\widehat{\mathcal{M}}_{\ssquare\ssquare}\!\left(A^{\otimes2}\right)
    ={}&
    \widehat{\mathcal{M}}_{\circ\circ}\!\left(A^{\otimes2}\right)
    -\widehat{\mathcal{M}}_{\ssquare\ssquare}\!\left(A^{\otimes2}\right)- 
    \Delta t\sum_{\alpha}
    [A, J_\alpha]\otimes[A,J_\alpha]
    +\mathcal{O}(\Delta t^{3/2}) \ .
\end{align}
Taking the continuous-time limit as in Eq.~\eqref{con_time_2replica}, the corresponding action of the off-diagonal generator is therefore
\begin{equation}\label{B_off_diag}
    \widehat{\mathcal{R}}_{\ssquare\circ}\!\left(A^{\otimes2}\right)
    =
    \widehat{\mathcal{R}}_{\circ\circ}\!\left(A^{\otimes2}\right)
    -
    \widehat{\mathcal{R}}_{\ssquare\ssquare}\!\left(A^{\otimes2}\right)
    -
    \sum_{\alpha}
    [A, J_\alpha]\otimes[A, J_\alpha] \ .
\end{equation}
Here, as for the discrete maps, the hat indicates the operator representation associated with the specific initial boundary structure considered above.

We now return to the second term of Eq.~\eqref{OTOC_cont2}. The quantity to be evaluated is still written in the folded replica representation as $
    (\blackssquare^{}_{t-s}|
    \mathcal{R}_{\ssquare\circ}
    |\bullet^{}_{s})$.
To evaluate it, we now use the operator representation derived above.
Since the generator acts at time $s$ on the evolved initial boundary, the operator $A$ appearing in Eq.~\eqref{B_off_diag} is replaced by $A_{\mathcal L}(s)$.
In this representation, contraction with the evolved final boundary amounts to closing the two system replicas with the swap permutation and with the two insertions of $B_{\mathcal L^\dagger}(t-s)$. 
Using the swap contraction, we have
\begin{align}
    \frac{1}{d_A}
    (\blackssquare^{}_{t-s}|
    \mathcal{R}_{\ssquare\circ}
    |\bullet^{}_{s})
    ={}&
    \frac{1}{d_A}
    (\blackssquare^{}_{t-s}|
    \left(
    \mathcal{R}_{\circ\circ}
    -
    \mathcal{R}_{\ssquare\ssquare}
    \right)
    |\bullet^{}_{s})-
    \sum_{\alpha}
    \left\langle
    [A_{\mathcal L}(s),J_\alpha]
    B_{\mathcal L^\dagger}(t-s)
    [A_{\mathcal L}(s),J_\alpha]
    B_{\mathcal L^\dagger}(t-s)
    \right\rangle \ .
    \label{eq:offdiag_split}
\end{align}
The first term on the right-hand side is a total derivative. Indeed,
defining
\begin{equation}
    f(s)
    =
    \frac{1}{d_A}
    (\blackssquare^{}_{t-s}|\bullet^{}_{s}) \ ,
\end{equation}
and using
$    |\bullet^{}_{s})
    =
    e^{s\mathcal{R}_{\circ\circ}}|\bullet)$ and $
    (\blackssquare^{}_{t-s}|
    =
    (\blackssquare|
    e^{(t-s)\mathcal{R}_{\ssquare\ssquare}} $,
one finds
\begin{equation}
    \frac{d}{ds}f(s)
    =
    \frac{1}{d_A}
    (\blackssquare^{}_{t-s}|
    \left(
    \mathcal{R}_{\circ\circ}
    -
    \mathcal{R}_{\ssquare\ssquare}
    \right)
    |\bullet^{}_{s}) \ .
    \label{eq:total_derivative}
\end{equation}
Substituting into Eq.~\eqref{OTOC_cont2} gives
\begin{align}
    \langle A(t)BA(t)B\rangle_{\mathcal{L}}
    ={}&
    f(0)
    +
    \int_0^t ds\,\frac{d}{ds}f(s)-
    \sum_{\alpha}
    \int_0^t ds\,
    \left\langle
    [A_{\mathcal L}(s),J_\alpha]
    B_{\mathcal L^\dagger}(t-s)
    [A_{\mathcal L}(s),J_\alpha]
    B_{\mathcal L^\dagger}(t-s)
    \right\rangle \ ,
\end{align}
where $f(0) =     \frac{1}{d_A}(\blackssquare^{}_{\,t}|\bullet) $ as in Eq.~\eqref{eq:supp_diagonal_otoc_2}.
The first two terms combine to give $f(t) = \frac{1}{d_A}
    (\blackssquare|\bullet^{}_{t})
    =
    \left\langle
    A_{\mathcal L}(t) B A_{\mathcal L}(t) B
    \right\rangle $.
Therefore,
\begin{align}
\left\langle A(t)BA(t)B\right\rangle_{\mathcal L}
={}&
\left\langle
A_{\mathcal L}(t)BA_{\mathcal L}(t)B
\right\rangle-
\sum_{\alpha}\int_0^t ds\,
\left\langle
[A_{\mathcal L}(s),J_\alpha]
B_{\mathcal L^\dagger}(t-s)
[A_{\mathcal L}(s),J_\alpha]
B_{\mathcal L^\dagger}(t-s)
\right\rangle \ .
\label{eq:supp_full_otoc}
\end{align}

This completes the derivation of the continuous-time OTOC from the two-replica dynamics of the minimal circuit.

\bigskip

Some comments are in order. \\ The single-copy dynamics discussed in the previous section can be understood as the $k=1$ instance of the same replica construction. In this case, there is only one possible non-crossing partition, namely the identity, so that the replica structure is trivial and the corresponding transfer matrix reduces to the single-copy channel $\mathcal{M}$.\\
More generally, Ref.~[\reffritzschfree] develops the replica construction for arbitrary replica number $k$. While the discrete replica construction is therefore available for arbitrary $k$, in this work we have derived its continuous-time limit explicitly for $k=1$ and $k=2$. In principle, the same procedure can be carried out at any order. Indeed, the operator representation introduced in Eq.~\eqref{M_unfolded} admits the direct generalization
\begin{equation}
    \widehat{\mathcal{M}}_{\nu\sigma}^{(k)}\!\left(A^{\otimes k}\right)
    =
    \frac{1}{d_C^k}
    \operatorname{Tr}_{C^{\otimes k}}
    \left[
        U^{\otimes k}
        \left(
            A^{\otimes k}\otimes P_\sigma^{-1}
        \right)
        U^{\dagger\otimes k}
        \left(
            \mathbb{1}_A^{\otimes k}\otimes P_\nu
        \right)
    \right] \ ,
\end{equation}
where $\nu$ and $\sigma$ label the corresponding non-crossing partitions, and $P_\nu$ and $P_\sigma$ act on the $k$ replicated copies of $C$. Starting from this expression, one can in principle expand each replica map in $\Delta t$ following the same procedure used above and derive the coupled continuous-time generators associated with the non-crossing-partition sectors which enter in the transfer matrix. Carrying out this construction explicitly for arbitrary $k$ would provide access to higher-order correlation functions and is left for future work. 

\subsection{Eigenoperator dynamics}
The above expressions simplify when considering $A$ and $B$ as right and left eigenoperators of the Lindbladian superoperator, $\mathcal{L}(A) = -\phi A, \, \mathcal L^\dagger(B)=-\phi B$. The correlation function simplifies as 
 \begin{equation}
     \langle A_\mathcal{L}(t)B\rangle = e^{-\phi t}\langle AB\rangle \ . 
 \end{equation}
This is the continuous-time limit expression of Eq.~(44) in Ref.~[\reffritzschfree]. 
For the OTOC,
 \begin{equation}
\langle A(t)BA(t)B\rangle_\mathcal{L} = e^{-2\phi t} \langle ABAB\rangle - t e^{-2\phi t}  \sum_{\alpha} \langle   \left[A,J_\alpha\right]B \left[A,J_\alpha\right]B\rangle\ .
 \label{formula_OTOC_eigenoperators}
 \end{equation}
This is the continuous-time limit expression of Eq.~(20) in Ref.~[\reffritzschcumulants].  As a consequence, for generic traceless local operators, the OTOC rate is governed by twice the smallest (in modulus) nonzero eigenvalue, with polynomial corrections. The latter ultimately come from the presence of Jordan blocks in the transfer matrix~\eqref{transfer_matrix}.\\

\section{Single-qubit example}
\label{supp_sec_onequbit}

As a simple illustrative example, we consider a single qubit undergoing unitary rotation on the Bloch sphere and subject to dephasing and depolarization. The formalism developed in our work allows us to compute analytically correlation functions and OTOCs, and to compare the latter with those from decoupled dynamics of the two replicas.
We fix the Hamiltonian and the jump operators 
\begin{equation}
    H = p \sigma_z\ ,\qquad \quad  J_\alpha =\sqrt{\gamma_\alpha}\sigma_\alpha ,\ \ \alpha=x,y,z\ ,
    \label{single_ham}
\end{equation}
where $\sigma_\alpha$ are Pauli operators. 
We focus on the case $\gamma_x=\gamma_y \equiv \gamma$.    
The resulting Lindbladian acts as
\begin{equation}
    \mathcal{L}( \rho ) = -ip [\sigma_z,\rho ] + \sum_{\alpha = x,y,z}\gamma_\alpha(\sigma_\alpha \rho \sigma_\alpha - \rho )    
\end{equation}
and can be diagonalized exactly. It has four eigenvalues $\phi_{\mathbb{1}}=0, \phi_z = \theta, \phi_{\pm}=\Gamma \pm i\omega$, respectively related to the dynamics of the right and left eigenoperators $\mathbb{1}$, $\sigma_z$, and $\sigma_\pm= \frac{1}{2}(\sigma_x \pm i\sigma_y)$; we defined $\theta=4\gamma$, $\Gamma=2\gamma+2\gamma_z$, and
$\omega=2p$.

\begin{figure}[t]
	\centering
	\includegraphics[width=.85\linewidth]{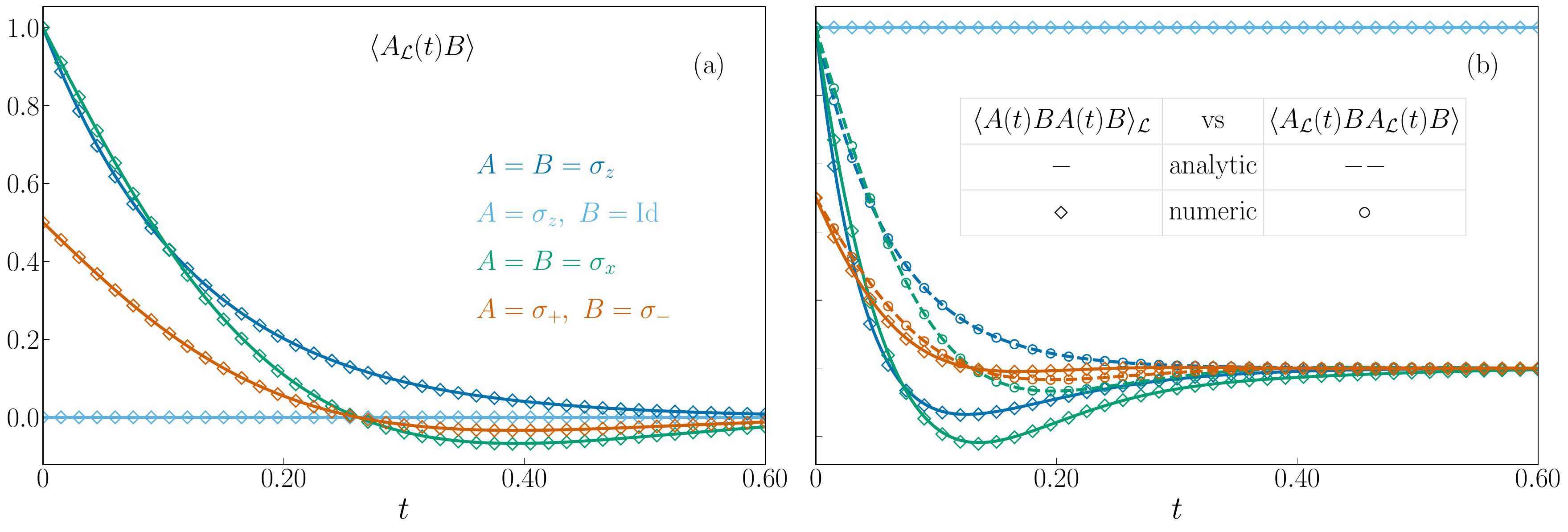}
	\caption{Single rotating qubit subject to dephasing and depolarization, cf.~Eq.~\eqref{single_ham}. The parameters are fixed to $p=3$, $\gamma_z=1$, and $\gamma=2$. We consider correlations between $A=B=\sigma_z$ (blue), $A=\sigma_z$ and $B=\mathbb{1}$ (light blue), $A=B=\sigma_x$ (green), and $A=\sigma_+$ and $B=\sigma_-$ (orange).
		(a) Correlation functions $\langle A_{\mathcal L}(t)B\rangle$. The analytical expressions (solid lines) are compared with numerical simulations of the Lindblad evolution (diamonds), using a time step $\Delta t=10^{-3}$.
		(b) Full OTOCs $\langle A(t)BA(t)B\rangle_{\mathcal L}$ (solid lines) compared with the corresponding reduced OTOCs $\langle A_{\mathcal L}(t)BA_{\mathcal L}(t)B\rangle$ (dashed lines).  
		The analytical results follow from Eqs.~\eqref{2ptcorr_single} and~\eqref{single_int}, with the latter giving the correction that distinguishes the full from the reduced OTOCs. Numerically, the full OTOC (diamonds) is obtained by simulating the coupled evolution of the two replicas, whereas the reduced OTOC (circles) is obtained from the single-copy Lindblad evolution and subsequently evaluating $\langle A_{\mathcal L}(t)BA_{\mathcal L}(t)B\rangle$.}
	\label{OneQubit}
\end{figure}

Correlators for its eigenoperators and consequently for Pauli matrices can be computed analytically. 
These are illustrated in Fig.~\ref{OneQubit}, which shows analytical curves together with numerical simulations.

\begin{itemize}
    \item For the correlation functions, for instance:
\begin{align}
\langle \sigma_z(t)\sigma_z\rangle_{\mathcal{L}} = e^{-\theta t}, \qquad 
\langle \sigma_x(t)\sigma_x\rangle_{\mathcal{L}} = e^{-\Gamma t} \cos{\omega t}, \qquad
\langle \sigma_+(t)\sigma_-\rangle_{\mathcal{L}} = \frac{1}{2}e^{-\Gamma t} (\cos{\omega t}-i\sin{\omega t}) \ .
\label{2ptcorr_single}
\end{align}
\item The corresponding reduced OTOCs exhibit a decay rate twice that of the correlation functions.
\item The replica-correlation term, expressed by the integral $I_{AB}(t)$, exhibits a polynomial prefactor multiplying the double-rate exponential decay, reflecting the presence of Jordan blocks in the coupled evolution, as mentioned already for Eq.~\eqref{formula_OTOC_eigenoperators}:
\begin{align}
    I_{zz}(t) = -2\theta t   e^{-2\theta t} , \quad
    I_{xx}(t) = -4\gamma_z t   e^{-2\Gamma t}\cos{2\omega t}  - \theta t  e^{-2\Gamma t} , \quad
    I_{+-}(t) = -2\gamma_z t e^{-2\Gamma t}(\cos{2\omega t}-i\sin{2\omega t})\  .
\label{single_int}
\end{align}
The major difference between the full and reduced OTOCs is therefore the nonmonotonic decay of the former stemming from $I_{AB}(t)$ and its additional linear prefactor.
\end{itemize}
Note that, for $B=\mathbb{1}$ and $A=\sigma_z$, the uncorrelated OTOC decays as $e^{-2\theta t}$, whereas the full OTOC remains equal to $\langle \sigma_z^2\rangle=1$. The replica contribution,
\begin{equation}
I_{z,\mathbb{1}}(t)=1-e^{-2\theta t},
\end{equation}
therefore exactly compensates the dissipative loss of Hilbert--Schmidt norm, illustrating the conservation of quantum information discussed in the main text.\\

\section{Additional Proofs}
\label{sec:proof}

\subsection{Multiplicative structure of the dynamics}

The conservation of the Hilbert--Schmidt norm discussed in the main text is a particular instance of a more general property. For general OTOCs of the form $\braket{A(t)BC(t)D}$, a direct extension of the derivation in Sec.~\ref{supp_sec_derivation} returns a full Lindblad evolution of the form
\begin{equation}\label{eq:OTOC_ABCD}
    \left\langle
    A(t)BC(t)D
    \right\rangle_{\mathcal L}
    = 
    \left\langle
    A_{\mathcal L}(t)BC_{\mathcal L}(t)D
    \right\rangle
    -\sum_\alpha\int_0^t ds\,
    \left\langle
    [A_{\mathcal L}(s),J_\alpha] B_{\mathcal L^\dagger}(t-s)
    [C_{\mathcal L}(s),J_\alpha]
    D_{\mathcal L^\dagger}(t-s)
    \right\rangle \ .
\end{equation}

Fixing $B=\mathbb{1}$, the closed-system OTOC trivially satisfies $\braket{A(t)C(t)D}=\braket{(AC)(t)D}$. This property is lost in the reduced OTOC, since for generic dynamics $\braket{A_{\mathcal{L}}(t)C_{\mathcal{L}}(t) D} \neq \braket{(AC)_{\mathcal{L}}(t)D}$. However, this multiplicative property is restored in the full OTOC of Eq.~\eqref{eq:OTOC_ABCD}. 
Let us consider the full correlator with $B=\mathbb{1}$, which we denote as
\begin{equation}
    F_{AC,D}(t)
    \equiv     \left\langle
    A(t)C(t)D
    \right\rangle_{\mathcal L}
    = 
    \left\langle
    A_{\mathcal L}(t)C_{\mathcal L}(t)D
    \right\rangle
    +
    I_{AC,D}(t)\ ,
\end{equation}
where the replica contribution is
\begin{equation}
    I_{AC,D}(t)
    =
    -\sum_\alpha\int_0^t ds\,
    \left\langle
    [A_{\mathcal L}(s),J_\alpha]
    [C_{\mathcal L}(s),J_\alpha]
    D_{\mathcal L^\dagger}(t-s)
    \right\rangle \ .
    \label{eq:I_general}
\end{equation}
For compactness, in the following we denote $A_t=A_{\mathcal L}(t)$ and $C_t=C_{\mathcal L}(t)$.

We first consider the time derivative of the first contribution
\begin{align}
    \frac{d}{dt}
    \left\langle A_t C_t D\right\rangle
     =
\left\langle
\mathcal L(A_t)C_tD
\right\rangle
+
\left\langle
A_t\mathcal L(C_t)D
\right\rangle = 
    \left\langle
    \mathcal L(A_tC_t)D
    \right\rangle
    +
    \sum_\alpha
    \left\langle
    [A_t, J_\alpha][C_t, J_\alpha]D
    \right\rangle \ .
    \label{eq:uncorr_product}
\end{align}
Here we used the following Leibniz identity, satisfied by the Lindblad generator for Hermitian jump operators
\begin{equation}
\mathcal L(AC)
=
\mathcal L(A)C
+
A\mathcal L(C)
-
\sum_\alpha[A,J_\alpha][C, J_\alpha] \ .
\end{equation}
The reduced OTOC does not preserve the multiplicative structure of the microscopic dynamics, precisely due to the additional term in Eq.~\eqref{eq:uncorr_product}.
We now show that the replica correction accounts for this discrepancy.

Differentiating the replica contribution in Eq.~\eqref{eq:I_general}, we obtain
\begin{align}
    \frac{d}{dt}I_{AC,D}(t)
    ={}&
    -\sum_\alpha
    \left\langle
    [A_t,J_\alpha]
    [C_t,J_\alpha]D
    \right\rangle
    -
    \sum_\alpha\int_0^t ds\,
    \langle
    [A_{\mathcal L}(s),J_\alpha]
    [C_{\mathcal L}(s),J_\alpha]
    \frac{d}{dt}D_{\mathcal L^\dagger}(t-s)
    \rangle \ ,
    \label{eq:I_derivative_1}
\end{align}
where 
\begin{equation}
\frac{d}{dt}D_{\mathcal L^\dagger}(t-s)=\mathcal L^\dagger \left(D_{\mathcal L^\dagger}(t-s)\right) =
    \mathcal L^\dagger
    e^{(t-s)\mathcal L^\dagger}(D)=
    e^{(t-s)\mathcal L^\dagger}
    \mathcal L^\dagger(D)= 
    \left(
    \mathcal L^\dagger D
    \right)_{\mathcal L^\dagger}(t-s)\ .
\end{equation}
The derivative of the full correlator is therefore
\begin{align}
    \frac{d}{dt}F_{AC,D}(t)
    ={}&
    \left\langle
    A_tC_t\mathcal L^\dagger (D)
    \right\rangle-
    \sum_\alpha\int_0^t ds\,
    \left\langle
    [A_{\mathcal L}(s),J_\alpha]
    [C_{\mathcal L}(s),J_\alpha]
    \left(\mathcal L^\dagger
    D\right)_{\mathcal L^\dagger}(t-s)
    \right\rangle \ ,
\end{align}
where we used the adjoint relation $\left\langle \mathcal{L}(X)D \right\rangle = \left\langle X \mathcal{L}^\dagger (D)\right\rangle$ for the first term. The second term is precisely
$I_{AC,\mathcal L^\dagger D}(t)$, so that
\begin{equation}
    \frac{d}{dt}F_{AC,D}(t)
    =
    F_{AC,\mathcal L^\dagger D}(t) \ .
    \label{eq:F_evolution}
\end{equation}
To identify the solution of this equation, consider
\begin{equation}
    G_{AC,D}(t)
    :=
    \left\langle
    (AC)_{\mathcal L}(t)D
    \right\rangle .
\end{equation}
Its time derivative is
\begin{align}
    \frac{d}{dt}G_{AC,D}(t)
    &=
    \left\langle
    \mathcal L\left((AC)_{\mathcal L}(t)\right)D
    \right\rangle = 
    \left\langle
    (AC)_{\mathcal L}(t)
    \mathcal L^\dagger(D)
    \right\rangle=
    G_{AC,\mathcal L^\dagger D}(t) \ .
\end{align}
Thus, $F_{AC,D}(t)$ and $G_{AC,D}(t)$ obey the same evolution equation.
Since furthermore $F_{AC,D}(0)=\langle ACD\rangle = G_{AC,D}(0)$, it follows that the two functions coincide
\begin{equation}
    F_{AC,D}(t)
    = G_{AC,D}(t) = 
    \left\langle
    (AC)_{\mathcal L}(t)D
    \right\rangle \qquad \longrightarrow \qquad 
\left\langle
    A(t)C(t)D
    \right\rangle_{\mathcal L}
    =
    \left\langle
    (AC)_{\mathcal L}(t)D
    \right\rangle \ .
    \label{eq:multiplicativity_restored}
\end{equation}
Equivalently, $
    I_{AC,D}(t)
    =
    \left\langle
    (AC)_{\mathcal L}(t)D
    \right\rangle
    -
    \left\langle
    A_{\mathcal L}(t)C_{\mathcal L}(t)D
    \right\rangle. $
This shows that the replica correction restores the multiplicative structure of the underlying unitary dynamics. This also has a direct consequence for the decay of the two contributions: in the reduced correlator, $A$ and $C$ evolve independently and their individual decay rates combine, whereas the full correlator is governed by the Lindblad evolution of the product $AC$, whose decay can in general be different.

\subsection{Late-time freeness}

We now consider the long-time limit of the correlation function and of the full OTOC. 
We aim to show that their asymptotic values satisfy
\begin{align}
\label{eq:freeness_app}
\lim_{t\rightarrow\infty}\langle A(t)B\rangle
&=
\langle A\rangle\langle B\rangle \ ,
\\
\lim_{t\rightarrow\infty}\langle A(t)BA(t)B\rangle
&=
\langle A^2\rangle\langle B\rangle^2
+\langle B^2\rangle\langle A\rangle^2
-\langle A\rangle^2\langle B\rangle^2 \ .
\nonumber
\end{align}

For ergodic Lindblad dynamics considered, the only non-decaying eigenoperator of the Lindbladian is the identity matrix satisfying $\mathcal{L}(\mathbb 1) = 0$, which is both a right and left eigenoperator. As such, the long-time evolution relaxes observables to their infinite-temperature expectation values,
\begin{equation}
\lim_{t\rightarrow\infty} A_{\mathcal L}(t)
=
\langle A\rangle \mathbb 1 \ ,
\qquad
\lim_{t\rightarrow\infty} B_{\mathcal L^\dagger}(t)
=
\langle B\rangle \mathbb 1 \ .
\end{equation}

As a consequence, for the correlation function one immediately finds
\begin{equation}
\lim_{t\rightarrow\infty}
\langle A_{\mathcal L}(t)B\rangle
=
\langle A\rangle\langle B\rangle \ .
\end{equation}

For the full OTOC,
\begin{align}
\langle A(t)BA(t)B\rangle_{\mathcal L}
={}&
\langle A_{\mathcal L}(t)BA_{\mathcal L}(t)B\rangle -
\sum_{\alpha}\int_0^t ds\,
\left\langle
[A_{\mathcal L}(s),J_\alpha]
B_{\mathcal L^\dagger}(t-s)
[A_{\mathcal L}(s),J_\alpha]
B_{\mathcal L^\dagger}(t-s)
\right\rangle \ ,
\end{align}
the first contribution has the long-time limit
\begin{equation}
\lim_{t\rightarrow\infty}
\langle
A_{\mathcal L}(t)BA_{\mathcal L}(t)B
\rangle
=
\langle A\rangle^2\langle B^2\rangle \ .
\end{equation}
To evaluate the replica-correlation term, we note that the integrand vanishes when
$s$ is of order $t$, since in this case 
$
[A_{\mathcal L}(s),J_\alpha]
\longrightarrow
\langle A\rangle[\mathbb 1,J_\alpha]
=
0
$.
For $s\ll t$, instead,
$
B_{\mathcal L^\dagger}(t-s)
\longrightarrow
\langle B\rangle\mathbb 1
$,
so that
\begin{align}
&-\lim_{t\rightarrow\infty}
\sum_\alpha \int_0^t ds\,
\left\langle
[A_{\mathcal L}(s),J_\alpha]
B_{\mathcal L^\dagger}(t-s)
[A_{\mathcal L}(s),J_\alpha]
B_{\mathcal L^\dagger}(t-s)
\right\rangle
=
-\langle B\rangle^2
\sum_\alpha\int_0^\infty ds\,
\left\langle
[A_{\mathcal L}(s),J_\alpha]^2
\right\rangle \ .
\end{align}
Using the identity
\begin{equation}
\frac{d}{ds}
\left\langle
A_{\mathcal L}(s)A_{\mathcal L}(s)
\right\rangle
=
\sum_\alpha
\left\langle
[A_{\mathcal L}(s),J_\alpha]^2
\right\rangle \ ,
\end{equation}
we obtain
\begin{align}
\sum_\alpha\int_0^\infty ds\,
\left\langle
[A_{\mathcal L}(s),J_\alpha]^2
\right\rangle
&=
\lim_{s\rightarrow\infty}
\left\langle
A_{\mathcal L}(s) A_{\mathcal L}(s)
\right\rangle
-
\left\langle
A^2
\right\rangle
=
\langle A\rangle^2-\langle A^2\rangle \ .
\end{align}
Therefore,
\begin{align}
\lim_{t\rightarrow\infty}
\langle A(t)BA(t)B\rangle_{\mathcal L}
&=
\langle A\rangle^2\langle B^2\rangle
-\langle A\rangle^2\langle B\rangle^2
+\langle A^2\rangle\langle B\rangle^2 \ ,
\end{align}
in agreement with Eq.~\eqref{eq:freeness_app}.

\end{document}